\documentclass[amsmath,amssymb,nofootinbib,a4paper,12pt]{article}
\pdfoutput=1
\usepackage{jheppub}

\usepackage{bm}
\usepackage{ulem}

\usepackage{amsmath,epsfig}
\usepackage{amssymb,amsfonts}
\usepackage{latexsym}
\usepackage[latin1]{inputenc}

\usepackage{tocvsec2}
\usepackage{subeqnarray}
\usepackage{xcolor}

\usepackage{graphicx}
\usepackage{longtable}

\relax
\def\be{\begin{equation}}
\def\ee{\end{equation}}
\def\bea{\begin{eqnarray}}
\def\eea{\end{eqnarray}}

\newcommand\fverb{\setbox\pippobox=\hbox\bgroup\verb}
\newcommand\fverbdo{\egroup\medskip\noindent%
                        \fbox{\unhbox\pippobox}\ }
\newcommand\fverbit{\egroup\item[\fbox{\unhbox\pippobox}]}

\newcommand{\bear}{\begin{eqnarray}}

\newcommand{\eear}{\end{eqnarray}}

\newcommand{\bsea}{\begin{subeqnarray}}
\newcommand{\esea}{\end{subeqnarray}}
\newbox\pippobox

\def\6{\partial}

\def\pa{\partial}

\def\m{\mu}
\def\n{\nu}

\def\sp{\;\;\;,\;\;\;}

\def\ts{ \tilde S }

\newcommand{\comments}[1]{}
\allowdisplaybreaks[3]

\begin{document}
\subheader{CCTP-2016-02\\CCQCN-2016-127\\ SU-ITP-16/03}

\renewcommand*{\thefootnote}{\fnsymbol{footnote}}
\title{Effective holographic theories of momentum relaxation and violation of conductivity bound}
\author[1,2]{Blaise Gout\'eraux,}
\affiliation[1]{Stanford Institute for Theoretical Physics, Department of Physics, Stanford University, Stanford, CA 94305-4060, USA}
\affiliation[2]{APC, Universit\'e Paris 7, CNRS/IN2P3, CEA/IRFU, Obs. de Paris, Sorbonne Paris Cit\'e, B\^atiment Condorcet, F-75205, Paris Cedex 13, France (UMR du CNRS 7164).}
\author[2,3]{Elias Kiritsis\footnote{\href{http://hep.physics.uoc.gr/~kiritsis/}{http://hep.physics.uoc.gr/~kiritsis/}}}
\affiliation[3]{Crete Center for Theoretical Physics and I.P.P.,
Department of Physics, University of Crete, 71003 Heraklion, Greece}
\author[3]{and~Wei-Jia Li}
\emailAdd{gouterau@stanford.edu}
\emailAdd{weijiali@physics.uoc.gr}
\abstract{We generalize current holographic models with homogeneous breaking of translation symmetry by incorporating higher derivative couplings, in the spirit of effective field theories. Focusing on charge transport, we specialize to two simple couplings between the charge and translation symmetry breaking sectors. We obtain analytical charged black brane solutions and compute their DC conductivity in terms of horizon data. We constrain the allowed values of the couplings and note that the DC conductivity can vanish at zero temperature for strong translation symmetry breaking, thus showing that in general there is no lower bound on the conductivity.}


\maketitle
\renewcommand*{\thefootnote}{\arabic{footnote}}	
\setcounter{footnote}{0}
\section{Introduction}
In the absence of spontaneously broken symmetries, momentum relaxation is necessary to obtain non-divergent DC conductivities at finite density as the electric current generally overlaps with the momentum current. If momentum is conserved, there is a $\delta$-function in the electric conductivity at zero frequency associated to translation invariance. In linearized relativistic hydrodynamics \cite{Kovtun:2012rj}, the electric conductivity reads
\begin{equation}
\sigma(\omega)=\frac{i}{\omega}G^R_{JJ}(\omega)=\sigma_Q+\frac{\rho^2}{(\epsilon+p)}\left(\pi\delta(\omega)+\frac{i}\omega\right)
\end{equation}
in terms of the energy, pressure and charge densities and where $\sigma_Q$ is the conductivity of the `incoherent' current orthogonal to momentum $J_{inc}=J-\rho P/(\epsilon+p)$ \cite{Davison:2015taa}. In the context of gauge/gravity duality, states at finite density and temperature are dual to charged black holes and the conductivity is computed by solving the coupled linearized Einstein-Maxwell equations with appropriate boundary conditions. For translation-invariant states like the Reissner-Nordstr\"om black hole, the result above is faithfully reproduced \cite{Hartnoll:2007ip}.

To broaden out the $\delta$-function and recover finite DC conductivities, translation symmetry must be broken. While this can be implemented holographically in full generality by considering inhomogeneous black holes and solving numerically partial differential equations \cite{Horowitz:2012ky,Hartnoll:2014cua,Donos:2014yya,Hartnoll:2015faa,Rangamani:2015hka,Langley:2015exa,Hartnoll:2015rza}, another approach involving spatially linear sources and homogeneous black hole backgrounds has proven analytically tractable and particularly fruitful \cite{Vegh:2013sk,Davison:2013jba,Blake:2013bqa,Davison:2013txa,Donos:2013eha,Andrade:2013gsa, Donos:2014uba,Gouteraux:2014hca,Donos:2014cya, Amoretti:2014zha, Davison:2014lua, Baggioli:2014roa, Davison:2015bea, Andrade:2015hpa}.\footnote{Remarkably, some analytical results can still be obtained for inhomogeneous spacetimes \cite{Hartnoll:2014cua,Donos:2014yya,Hartnoll:2015faa,Rangamani:2015hka,Donos:2015gia,Banks:2015wha,Donos:2015bxe,Hartnoll:2015rza}.} Two regimes are physically relevant. First, the coherent regime at slow momentum relaxation, where the optical conductivity displays a sharp peak, the width of which is set by the momentum decay rate $\Gamma\ll T$ where $T$ is the temperature. Operationally, this means there is a single purely imaginary pole in the lower half plane close to the real axis, governing the dynamics at long time scales or low frequencies.  $\Gamma$ can be calculated perturbatively within the memory matrix approach \cite{Hartnoll:2012rj,Mahajan:2013cja, Davison:2013txa,Lucas:2015pxa} or holographically \cite{Davison:2013jba, Davison:2014lua,Lucas:2015vna, Davison:2015bea, Blake:2015epa} and is related to the imaginary part of the two-point function of the operator breaking translations. Consequently, the dynamics is not universal in this regime, as it depends on the scheme of translation symmetry breaking. Effective theories based on hydrodynamics with almost conserved momentum have been proposed \cite{Hartnoll:2007ih} and provide a good match to holographic calculations where applicable \cite{Davison:2014lua,Davison:2015bea, Blake:2015epa}.\footnote{See also \cite{Lucas:2015lna} for hydrodynamics around an inhomogeneous equilibrium state.}

The second regime of interest is the incoherent regime, when momentum relaxes quickly and is dominated by diffusion \cite{Hartnoll:2014lpa}. Accordingly, there is no longer an isolated pole close to the real axis and the optical conductivity is approximately constant at low frequencies. The crossover between the two regimes can be precisely captured by holographic computations \cite{Kim:2014bza, Davison:2014lua}. The DC conductivity obeys an Einstein relation and is set by the diffusion constant. Therefore, it is susceptible to be universal and can perhaps obey bounds.

A celebrated example of such a bound is the ratio of the shear viscosity to entropy density in strongly coupled quantum field theories. The bound can be computed holographically \cite{Kovtun:2004de}, though it is sensitive to higher derivative corrections \cite{Brigante:2007nu,Brigante:2008gz} (see \cite{Cremonini:2011iq} for a review) or translation symmetry breaking \cite{Jain:2014vka,Davison:2014lua,Jain:2015txa,Hartnoll:2016tri, Alberte:2016xja}. More recently, the possibility of a bound on the conductivity in the incoherent regime was conjectured to be responsible for the linear temperature dependence of many metallic materials \cite{Hartnoll:2014lpa}. It is known by now that insulating behavior is possible in holographic models at finite density, either because momentum relaxation becomes relevant in the IR \cite{Donos:2012js,Donos:2014uba,Gouteraux:2014hca,Donos:2014oha} or due to strong dynamics turning the system into a Mott-like insulator, \cite{Kiritsis:2015oxa,Ling:2015epa,Ling:2015exa}, or to the formation of charge density waves \cite{Ling:2014saa}.  It was recently shown however, that simple holographic models in four spacetime dimensions could not acccomodate a disorder-driven metal to insulator phase transition\footnote{Which should be understood as meaning the conductivity vanishes at fixed temperature as the strength of disorder is increased.} due to the existence of a lower bound on the conductivity \cite{Grozdanov:2015qia}.

The objectives of this paper are two-fold. First, in the spirit of effective holographic low energy theories, \cite{Charmousis:2010zz, Gouteraux:2011ce}, we would like to incorporate higher order terms involving the spatially linear sources responsible for momentum relaxation. These terms are allowed by symmetry and can either couple the sources to each other \cite{Baggioli:2014roa, Alberte:2015isw} (or to the gravity sector, though we will not consider this explicitly for simplicity) or to the charged sector. The latter possibility has not been studied until now. Second, we will compute the electric conductivity in two simple examples and study how the higher-derivative couplings affect the conductivity and the bound of \cite{Grozdanov:2015qia}.

The plan of the paper is as follows. In section \ref{section1}, we expand on the general philosophy of constructing effective holographic theories of momentum relaxation and write down a general action with higher derivative couplings between the translation symmetry breaking and charge degrees of freedom. In section \ref{section2} and \ref{section3}, we examine more closely two specific higher derivative couplings between the two sectors, find analytical black brane solutions parameterized by their temperature, chemical potential, and strength of translation symmetry breaking, and compute their DC conductivity. Finally, we discuss our results and conclude in section \ref{section4}. The covariant equations of motion are presented in appendix \ref{section5}. In appendices \ref{app:schroedcase2} and \ref{app:schroedcase3}, we derive constraints on the couplings at zero density.

\paragraph{Note added:} As this paper was being finalized, \cite{Baggioli:2016oqk} appeared which contains some overlap with our results. 

\section{General holographic effective theories of momentum relaxation \label{section1}}

We will construct here general effective actions for momentum relaxation in the limit in which the effects are smeared over all distances.
The proper way of describing these effects is in terms of the Stuckelberg mechanism for gravity whose Goldstone modes are the d scalars $X^{I}$ associated with the d spatial translations.

A covariant combination is given by the induced metric
\begin{equation}\label{1}
\hat g_{\m\n}=\frac{1}{d}\partial_{\m}X^{I}\partial_{\n}X^{I}.
\end{equation}
We further construct the mixed tensor as
\be
{X^{\m}}_{\n}=g^{\m\tau}\hat g_{\tau\n}
\label{2}\ee
and higher order tensors
\be
{({X^n})^{\mu}}_{\nu}={X^\mu}_{\alpha_1}{X^{\alpha_1}}_{\alpha_2}....{X^{\alpha_{n-1}}}_{\nu}
\label{2a}\ee
so that traces of the tensors like $Tr[X^n]\equiv{(X^n)^\mu}_\mu$
are scalars. We will include these  in Einstein-Maxwell-Dilaton (EMD) theories
\be
S=S_{R}+S_{\phi}+S_A\sp S_{R}=M^{d}\int d^{d+2}x\sqrt{-g}~R
\label{3}\ee
 \be
 S_{\phi}=M^{d}\int d^{d+2}x\sqrt{-g}\left[-{1\over 2}(\pa\phi)^2+Y(\phi)\right]\sp S_A=M^{d}\int d^{d+2}x\sqrt{-g}\left[-{Z(\phi)\over 4} F^2\right].
\label{4}\ee
To introduce momentum relaxation we can add couplings of $X$ to $S$ so that the theory is regular in the limit of $X\to 0$.
We define\footnote{In general we can add also derivatives of $X$, that would amount generically to  higher order equations of motion. This procedure becomes however uncontrollable and requires also higher derivative terms for the rest of the fields. We will therefore adopt a DBI ansatz in which we add $X$ but not its derivatives. Indeed the type of solutions we are interested in justify this kind of approximation.}
\be
\tilde S=\tilde S_{R}+\ts_{\phi}+\ts_{A}
\label{5}\ee
\be
\ts_R=M^{d}\int d^{d+2}x\sqrt{-g}~f(\phi,X)~R\sp
f(\phi,X)\equiv1+\sum_{n=1}^{\infty}f_n(\phi)Tr[X^n]
\label{6}\ee
\be
\ts_{\phi}=M^{d}\int d^{d+2}x\sqrt{-g}\left[-{1\over 2}G(\phi,X)(\pa\phi)^2+Y(\phi,X)\right],
\label{7}\ee
\be
 G(\phi,X)\equiv 1+
\sum_{n=1}^{\infty}G_n(\phi)Tr[X^n]\,,\qquad Y(\phi,X)\equiv
\sum_{n=0}^{\infty}Y_n(\phi)Tr[X^n]
\label{8}\ee
\be
\ts_A=-{M^d\over 4}\int d^{d+2}x\sqrt{-g}~ \left(\sum_{n,m=0}^{\infty} Z_{n,m}(\phi)Tr[X^m]Tr[X^nF^2]\right),
\label{9}\ee
where\footnote{The gauge related couplings are not the most general ones. The most general coupling includes traces of the type $Tr[X^m F X^n ~F]$. We will not consider these cases further in this paper. Also we could add multi trace terms. However for the type of backgrounds we consider such terms do not offer any novelties.} $Tr[X^nF^2]\equiv{X^{\mu}}_{\nu_1}...{X^{\nu_{n-1}}}_{\nu_n}{F^{\nu_n}}_{\nu_{n+1}}{F^{\nu_{n+1}}}_{\mu}$.

Our goal is to study how the new couplings affect the dynamics, both from the point of view of the background solution and of the transport of charge.

To proceed adiabatically, we will consider here the case without the scalar $\phi$. The effective action in this case is obtained from  (\ref{5})-(\ref{9}) by taking all functions of $\phi$ to be $\phi$
-independent.

\be
\tilde S=\tilde S_{R}+\ts_{A}\,,
\label{10}\ee
\be\ts_R=M^{d}\int d^{d+2}x\sqrt{-g}~\left[\left(1+\sum_{n=1}^{\infty}f_nTr[X^n]\right)~R+d(d+1)-
V(X)\right],
\label{11}\ee
\be
\ts_A=-{M^d\over 4}\int d^{d+2}x\sqrt{-g}~ \left(\sum_{n,m=0}^{\infty} Z_{n,m}~Tr[X^m]Tr[X^nF^2]\right),
\label{12}\ee
where $f_n$, $Z_{n,m}$ are all constant coefficients and $V(X)$ is a general function of $Tr[X^n]$.\footnote{In some previous work, the case of $V(Tr[X],det[X])$ has been considered, \cite{Alberte:2015isw}. For arbitrary $d>2$, our setup here is more general,  as $det[X]$ can be expressed in terms of $Tr[X^n]$. For $d=2$, it is equivalent.}  Here, we have set the AdS radius $L=1$ for simplicity.

In the following, we are going to focus on two examples that have not yet been studied in the literature, keeping $V(X)$ general for now: the axions $X^I$  couple directly to the kinetic term of the U(1) gauge boson either with
\begin{enumerate}
\item $Z_{1,0}\equiv \mathcal{J}$ constant, all other nontrivial terms vanishing.
\item  $Z_{0,1}\equiv -\mathcal{K}$ constant, all other nontrivial terms vanishing.
\end{enumerate}

We also choose a diagonal, rotationally symmetric ansatz
\be
ds^2=-D(r)dt^2+B(r)dr^2+C(r)dx^idx_i\sp A_t=A_t(r),
\label{13}\ee
and the scalars as
\be
X^I=k~\delta^I_{i}x^i,
\label{14}\ee
where $i$ denotes spatial directions. We derive the effective equations for the functions $B,C,D,A_t$. With the ansatz above, we will focus only on the case of $d=2$ (two spatial boundary directions), namely taking $i=x,y$.

\section{Example 1: General $V(X)$ and the $Tr[XF^2]$ coupling\label{section2}}
\subsection{Background solution}
In this case,  the action is simplified as
\begin{eqnarray}\label{15}
\tilde S=M^2\int d^{4}x\sqrt{-g}~\left[R+
6-V(X)-\frac{1}{4}F_{\mu\nu}F^{\mu\nu}-\frac{\mathcal{J}}{4}Tr[XF^2]\right],
\end{eqnarray}
For simplicity, we set the Planck scale  $M=1$ in appropriate units. With the ansatz (\ref{13}) and (\ref{14}), the equations of motion for $A_t$, $B$, $C$, $D$ are given by
\begin{eqnarray}\label{16}
\Big(\frac{C}{\sqrt{BD}}A_t'\Big)'=0\,,
\end{eqnarray}
\begin{eqnarray}\label{17}
&&\left[6-V\left(\frac{k^2}{C}\right)\right] BD+\frac{B'C'D}{BC}+\frac{1}{2}\frac{C'^2D}{C^2}-2\frac{C''D}{C}
-\frac{1}{2}A_t'^2=0\,,
\end{eqnarray}
\begin{eqnarray}\label{18}
&&\left[V\left(\frac{k^2}{C}\right)-6\right]  BD+\frac{C'D'}{C}+\frac{1}{2}\frac{C'^2D}{C^2}+\frac{1}{2}A_t'^2=0\,,
\end{eqnarray}
\begin{eqnarray}\label{19}
&&\left[V\left(\frac{k^2}{C}\right)-6\right]  BD-2\sum_{n=1}^\infty nV_{X^n}\left(\frac{k^2}{2C}\right)^{n}BD+\frac{1}{2}\left(\frac{C'D'}{C}-\frac{B'C'D}{BC}\right)-\frac{1}{2}\frac{C'^2D}{C^2}\nonumber\\
&&+\frac{C''D}{C}+D''-\frac{1}{2}\left(\frac{B'D'}{B}+\frac{D'^2}{D}+A_t'^2\right)=0\,,
\end{eqnarray}
where $V_{X^n}\equiv\frac{\partial V(X)}{\partial Tr[X^n]}$.  $\mathcal{J}$ does not appear in the background equations of motion, a consequence of $Tr[XF^2]=0$ for our choice of Ansatz. This implies that turning on this coupling does not affect the background solution and we expect to recover the solution of \cite{Bardoux:2012aw, Andrade:2013gsa}. We choose coordinates so that $C(r)=r^2$. Then,  from the
Maxwell equation (\ref{16}), we find
\begin{eqnarray}\label{20}
r^2A_t'=\rho\,,
\end{eqnarray}
where the constant $\rho$ is the charge density of the boundary system. Combining (\ref{17})and (\ref{18}), we obtain $B\propto D^{-1}$. The proportionality coefficient can be normalized by rescaling the coordinates.
The background solution is therefore given by
\begin{eqnarray}\label{21}
&&D(r)=\frac{1}{2r}\int_{r_h}^rds\left[6s^2-V\left(\frac{k^2}{s^2}\right)s^2\right] +\frac{1}{4}\Big(\frac{\rho^2}{r^{2}}-\frac{\rho^2}{r_hr}\Big)=B^{-1}(r)\,, \\ \nonumber
&&C(r)=r^2\,, \qquad  A_t(r)=\mu-\frac{\rho}{r},
\end{eqnarray}
where $r_h$ denotes the location of the event horizon (the outermost root of the function $D(r)$) and $\mu$ is the chemical potential. Requiring the gauge field to be regular on the horizon implies $\mu=\rho/r_h$. Finally, the Hawking temperature is
\begin{eqnarray}\label{22}
T=\frac{D'(r_h)}{4\pi}=\frac{1}{16\pi}\left[2r_h \left(6-V\left(\frac{k^2}{r_h^2}\right)\right) -\frac{\mu^2}{r_h}\right].
\end{eqnarray}
It is easy to check that the temperature vanishes at $r_h\neq0$ and $\mu\neq0$, so there is a non-zero extremal horizon radius and the IR geometry is AdS$_2\times R^2$.

While the background solution is independent of $\mathcal{J}$, the higher-derivative coupling will affect the linearized perturbations around the background, and hence charge transport properties as well. We analyze this next.

\subsection{DC Conductivity}

We now turn on small fluctuations around the background solution.
We denote $g_{\mu\nu}=\bar{g}_{\mu\nu}+\delta g_{\mu\nu}$, $A_\mu=\bar{A}_{\mu}+\delta A_{\mu}$ and $X^I=\bar{X}^I+\delta X^I$,
where the quantities with bars are evaluated on the background, and introduce the time-dependent perturbations as follows
\begin{eqnarray}
\delta A_\mu(t,r,x^i)=\int ^{+\infty}_{-\infty}\frac{d\omega d^{d-1}p_i}{(2\pi)^d}e^{-i \omega t+ip_ix^i}a_\mu(r),\\ \nonumber
\delta g_{\mu \nu}(t,r,x^i)=\int ^{+\infty}_{-\infty}\frac{d\omega d^{d-1}p_i}{(2\pi)^d}e^{-i \omega t+ip_ix^i}r^2h_{\mu \nu}(r), \\ \nonumber
\delta X^I(t,r,x^i)=\int ^{+\infty}_{-\infty}\frac{d\omega d^{d-1}p_i}{(2\pi)^d}e^{-i \omega t+ip_ix^i}\delta^I_i\psi^i(r).
\end{eqnarray}\label{23}
To derive the conductivity of the boundary system, we focus on homogeneous vector modes, setting all the momenta $p_i=0$. This means we only need to consider the $x$-component of the vector modes, namely $a_x$, $h_{tx}$, $h_{rx}$ and $\psi^{x}$.
The linearized Maxwell, scalar equation and Einstein equations read
\begin{eqnarray}
&&\left[\left(1-\frac{\mathcal{J}k^2}{4r^2}\right)D a_x'\right]'+\left(1-\frac{\mathcal{J}k^2}{4r^2}\right)\frac{\omega^2 }{D}a_x+\rho\left(h_{tx}'+i\omega h_{rx}\right)-\frac{\mathcal{J}\rho}{4}\Big[\left(\frac{k^2h_{tx}}{r^2}\right)'\nonumber\\
&&
+ik \omega\left(\frac{\psi^x}{r^2}\right)'-\frac{i k\omega}{r^2}({\psi^x}'-kh_{rx})\Big]=0\,,\\ \label{24a}
&&r^2\left(h_{tx}'+i\omega h_{rx}\right)+\frac{\rho}{r^{2}} a_x +\frac{ i k \bar{V}D}{\omega} ({\psi^x}'-kh_{rx})-\frac{\mathcal{J}}{4}\Big[\frac{kD\rho^2}{i \omega r^{4}}({\psi^x}'-kh_{rx})\\
\nonumber
&&+\frac{k^2\rho}{r^{4}} a_x\Big]=0\,,\\ \label{24b}
&&\left[r^{2} \bar{V} D({\psi^x}'-kh_{rx})\right]'+\frac{r^2\bar{V}}{D}\omega^2\psi^x-\frac{ikr^2 \bar{V}}{D}\omega h_{tx}+\frac{\mathcal{J}}{4}\Big\{\rho^2\Big[\Big(\frac{D({\psi^x}'-kh_{rx})}{r^2}\Big)'\nonumber\\ &&+\frac{\omega^2\psi^x-ik\omega h_{tx}}{Dr^2}\Big]-\frac{2ik \omega}{r^3}\rho a_x \Big \}=0\,,\\ \label{24c}
&&\frac{1}{2}M\left(k^2h_{tx}+ik\omega\psi^x\right)-\frac{D\left[r^{4}\left(h_{tx}'+i\omega h_{rx}\right)\right]'}{2r^2}-\frac{\rho}{2r^2}(1-\frac{\mathcal{J}k^2}{4r^2})D{a_x}'=0\,,\label{24d}
\end{eqnarray}
where
\begin{eqnarray}\label{25}
\bar{V}(r)\equiv\sum_{n=1}^\infty nV_{X^n}\left(\frac{Tr[X^n]}{2}\right)^{n-1}\Big|_{Tr[X^n]=\frac{k^{2n}}{r^{2n}}}
\end{eqnarray}
and $M(r)\equiv \bar{V}(r)+\frac{\mathcal{J}\rho^2}{4r^{4}}$. Absence of ghosts constrains $\bar{V}(r)$ to be always positive in the bulk \cite{Baggioli:2014roa}. To obtain the DC conductivity, we employ the Donos-Gauntlett strategy \cite{Donos:2014uba} and, skipping details, we arrive at the formula
\begin{eqnarray}\label{26}
\sigma_{DC}=\left(1-\frac{\mathcal{J}k^2}{4r_h^2}\right)+\left(1-\frac{\mathcal{J}k^2}{4r_h^2}\right)^2\frac{\mu^2 }{k^2 M(r_h)}\,.
\end{eqnarray}
For simplicity, we choose $V(X)=Tr[X]$ (see [22] for a discussion of the effect of arbitrary $V(X)$ on the conductivity), which leads to
\begin{eqnarray}\label{27}
M(r)=1+\frac{\mathcal{J}\rho^2}{4r^{4}}\,.
\end{eqnarray}

\begin{figure}
\begin{tabular}{cc}
\includegraphics[width=0.47\textwidth]{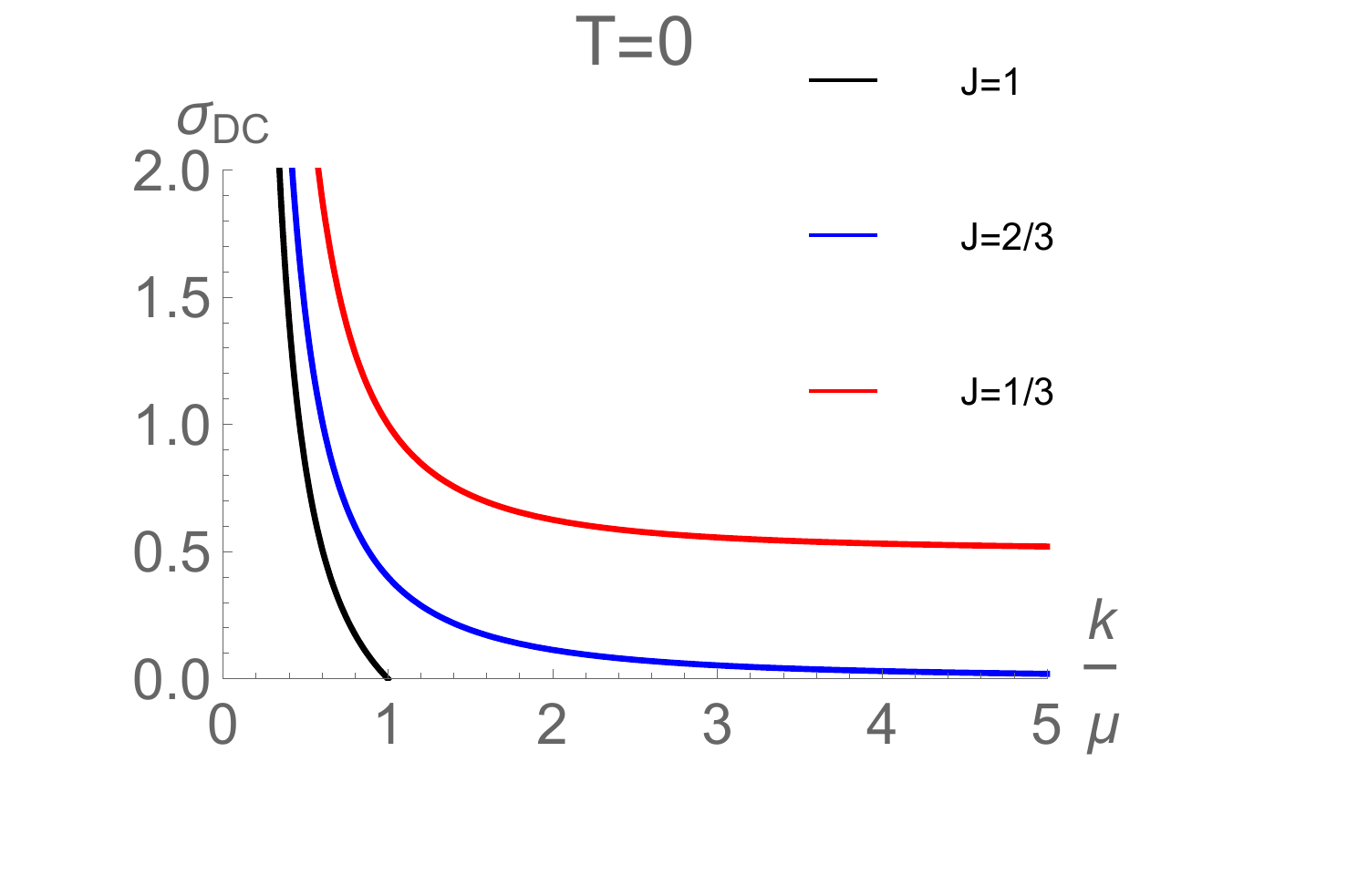} & \includegraphics[width=0.47\textwidth]{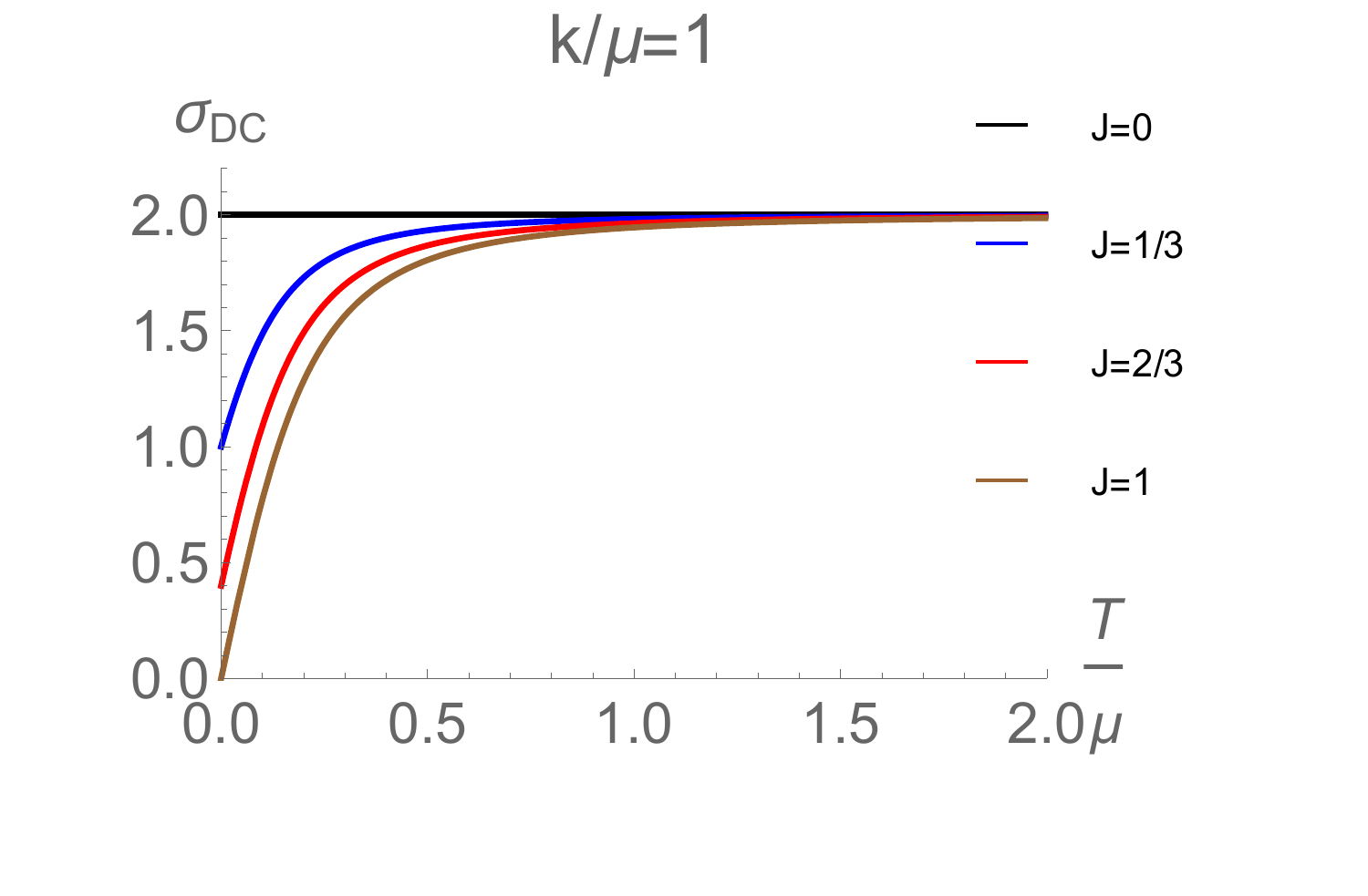}
\end{tabular}
\caption{Plots of the DC conductivity with the $\mathcal J$ coupling turned on. Left: We fix $T=0$ and vary $k/\mu$. Right: We fix $k=\mu$ and vary $T/\mu$. For $\mathcal J>2/3$, the DC conductivity vanishes at $T=0$ for $k=k_\textrm{max}$.}
\label{fig:1}
\end{figure}

\begin{figure}
\begin{center}{c}
\includegraphics[width=0.47\textwidth]{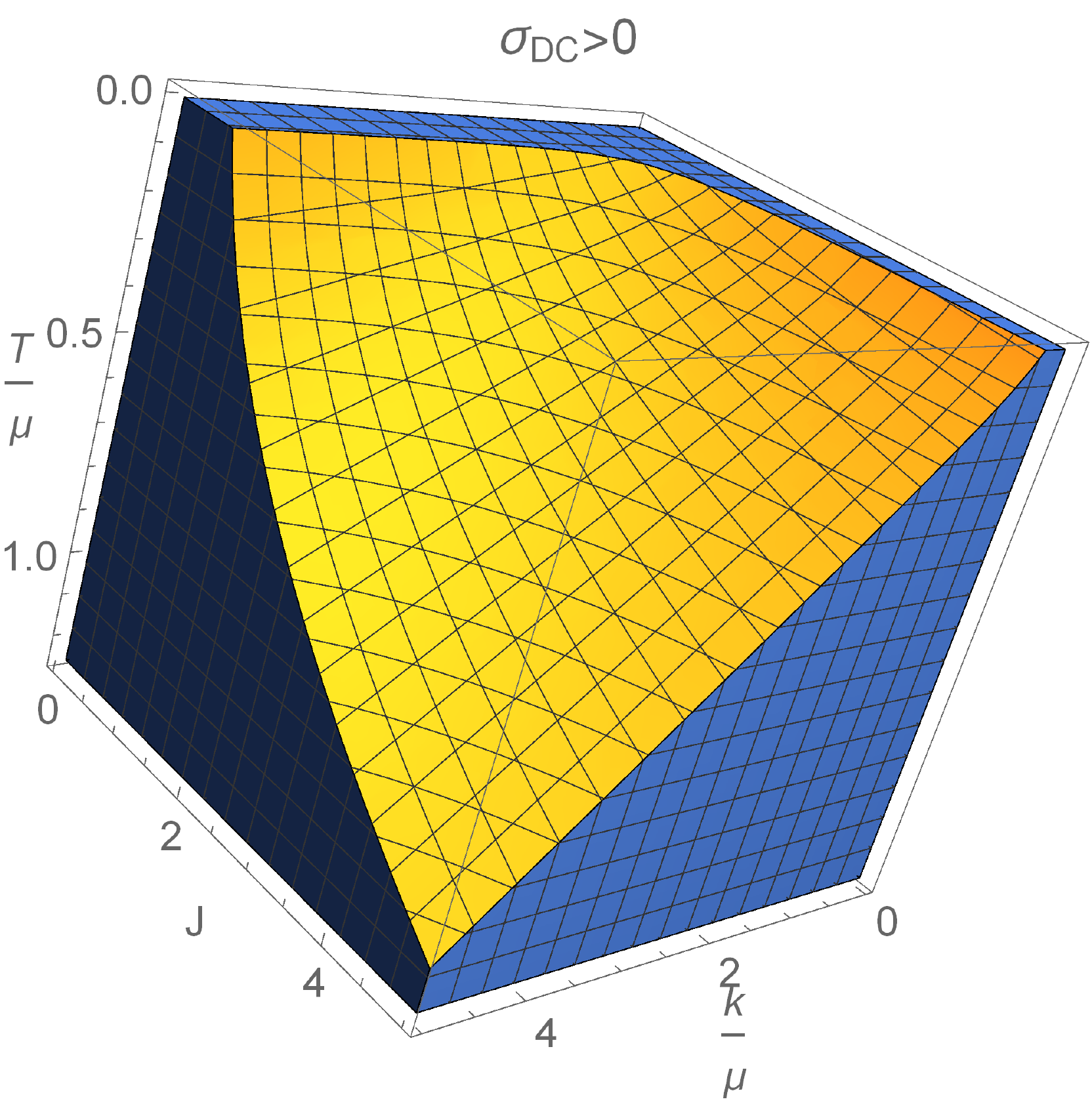}
\end{center}
\caption{Plot of the parameter range where the DC conductivity \eqref{26} is positive.}
\label{fig:sigmaDC23D}
\end{figure}

We start by analyzing the DC conductivity at zero temperature (but $\mu\neq0$, so the black hole is extremal), to investigate the presence of transitions from metallic to insulating behaviour (by which we mean that the DC conductivity vanishes at zero temperature) as a function of translation-symmetry breaking. Previous examples of metal to insulator transitions were studied in \cite{Donos:2012js,Donos:2014uba,Gouteraux:2014hca,Donos:2014oha}.
At zero temperature, the DC conductivity takes the simple form
\begin{equation}
\sigma_{DC}(T=0)=\frac{(\mu^2+k^2)\left(\mu^2+k^2(2-3\mathcal{J})\right)}{k^2\left(2k^2+\mu^2(1+3\mathcal{J})\right)}\,.
\end{equation}
This quantity is bounded from below by $1-3\mathcal{J}/2$ provided $\mathcal{J}<2/3$. However, for $\mathcal J>2/3$, the DC conductivity will inevitably vanish at $k_\textrm{max}^2=1/(3\mathcal J-2)$, and then become negative. At exactly $\mathcal J=2/3$, the DC conductivity vanishes only asymptotically. This is depicted in figure \ref{fig:1}. Note that for $\mathcal J<0$, the DC conductivity at small enough $k$ is negative.\footnote{Contrarily to the zero density case where $\mathcal J<0$ is allowed, see appendix \ref{app:schroedcase2}.}  Moreover, we show in appendix \ref{app:schroedcase2} that an instability develops for $\mathcal J>2/3$. So in the end we restrict the analysis to $0\leq\mathcal J\leq 2/3$ as causality requires the conductivity to be positive in general. Turning on temperature produces the same qualitative behaviour. In particular, the DC conductivity is still monotonous and asymptotes to
\begin{equation}
\sigma_{DC}\left(\frac{T}\mu,\frac{k}\mu\right)\underset{k/\mu\to+\infty}{\longrightarrow}1-\frac{3}2\mathcal J\,.
\end{equation}
In figure \ref{fig:sigmaDC23D}, we show the parameter range in which the DC conductivity is positive, as a function of $(T,k,\mathcal J)$.

\section{Example 2: General $V(X)$ and the $Tr[X] F^2$ coupling\label{section3}}

\subsection{Background Solution}
In this case, the action is given by
\begin{eqnarray}\label{28}
\tilde S=M^2 \int d^{4}x\sqrt{-g}~\left[R+
6-V(X)-\frac{1}{4}(1+\mathcal{K}Tr[X])F_{\mu\nu}F^{\mu\nu}\right],
\end{eqnarray}
and the EOMs for $A_t$, $B$, $C$ and $D$ are expressed as
\begin{eqnarray}\label{29}
\Big[\Big(1+\mathcal{K}\frac{k^2}{C}\Big)\frac{C}{\sqrt{BD}}A_t'\Big]'=0\,,
\end{eqnarray}
\begin{eqnarray}\label{30}
&&\left[6-V\left(\frac{k^2}{C}\right)\right] BD+\frac{B'C'D}{BC}+\frac{1}{2}\frac{C'^2D}{C^2}-2\frac{C''D}{C}-\frac{1}{2}\Big(1+\mathcal{K}\frac{k^2}{C}\Big)A_t'^2=0\,,
\end{eqnarray}
\begin{eqnarray}\label{31}
&&\left[V\left(\frac{k^2}{C}\right)-6\right]BD+\frac{C'D'}{C}+ \frac{1}{2}\frac{C'^2D}{C^2}+\frac{1}{2}\Big(1+\mathcal{K}\frac{k^2}{C}\Big)A_t'^2=0\,,
\end{eqnarray}
\begin{eqnarray}\label{32}
&&\left[V\left(\frac{k^2}{C}\right)-6\right]BD-2\sum_{n=1}^\infty nV_{X^n}\left(\frac{k^2}{2C}\right)^{n}BD+\frac{1}{2}\left(\frac{C'D'}{C}-\frac{B'C'D}{BC}\right)-\frac{1}{2}\frac{C'^2D}{C^2}\nonumber\\
&&+\frac{C''D}{C}+D''-\frac{1}{2}\left[\frac{B'D'}{B}+\frac{D'^2}{D}+ A_t'^2\right]=0\,.
\end{eqnarray}
Choosing the gauge $C(r)=r^2$, we again obtain $B=D^{-1}$ from (\ref{30}) and (\ref{31}). Inserting this into (\ref{29}) and solving the equation,  the  conserved charge now becomes
\begin{eqnarray}\label{33}
(r^2+\mathcal{K}k^2 )A_t'\equiv \rho\,.
\end{eqnarray}
Then the solution to the Maxwell field is given by
\begin{eqnarray}\label{34}
A_t(r)=\mu-\frac{\pi \rho}{2 k\sqrt{\mathcal K}}+\rho\frac{{_2}F_1(1,-\frac{1}{2};\frac{1}{2};-\frac{r^2}{\mathcal{K}k^2})-1}{r}
\end{eqnarray}
where ${_2}F_1$ is the hypergeometric function and $\rho$, $\mu$ are the charge density and chemical potential. We also find
\begin{eqnarray}\label{35}
D(r)=\frac{1}{2r}\int_{r_h}^rds\left[\left(6-V\left(\frac{k^2}{s^2}\right)\right)s^2-\frac{1}{2}\frac{\rho^2}{s^2+\mathcal{K}k^2}\right]=B^{-1}(r)\,.
\end{eqnarray}
Unlike the previous case,  the background solutions now depend on the coupling $\mathcal{K}$.\footnote{It would be interesting to work out how the thermodynamics are affected by $\mathcal K$ and whether this places constraints on the values it can take. We leave this for future work.}  The Hawking temperature is obtained as
\begin{eqnarray}\label{36}
T=\frac{1}{16\pi}\left[2r_h \left(6-V\left(\frac{k^2}{r_h^2}\right)\right)-\frac{\rho^2}{r_h^{3}\left(1+\mathcal{K}(k^2/{r_h}^2)\right)}\right].
\end{eqnarray}
From inspecting the zeroes of the temperature, it is easy to check that it vanishes at $\mu\neq0$ and $r_h\neq0$, so the IR geometry is AdS$_2\times R^2$ in spite of the non-trivial effect of the coupling $\mathcal K$ on the background geometry.

\subsection{DC Conductivity}

We now  study the fluctuations. In this case,  the linearized equations are
\begin{eqnarray}
&&\left[\left(1+\frac{\mathcal{K}k^2}{r^2}\right) D a_x'\right]'+\left(1+\frac{\mathcal{K}k^2}{r^2}\right)\frac{\omega^2}{D} a_x+\rho\left(h_{tx}'+i\omega h_{rx}\right)=0\,,\\ \label{37a}
&&r^2\left(h_{tx}'+i\omega h_{rx}\right)+\frac{\rho}{r^{2}} a_x +\frac{ i k D \bar{U}}{\omega} \left({\psi^x}'-k h_{rx}\right)=0\,,\\ \label{37b}
&&\left(r^{2} D \bar{U} {\psi^x}'\right)'+\frac{r^2 \bar{U} }{D}\omega^2\psi^x-k\left(r^{2} D \bar{U}h_{rx}\right)'-\frac{ikr^2 \bar{U}}{D}\omega h_{tx}=0\,,\\ \label{37c}
&&\frac{1}{2} \left(k^2\bar{U}h_{tx}+ik\bar{U}\omega\psi^x\right)-\frac{D\left[r^{4}\left(h_{tx}'+i\omega h_{rx}\right)\right]'}{2r^2}-\frac{\rho}{2r^2}D{a_x}'=0\,, \label{37d}
\end{eqnarray}
where $ \bar{U}(r)\equiv \bar{V}(r) -\frac{\mathcal{K}\rho^2}{2r^{4}[1+\mathcal{K}(k^2/r^2)]^2}$. Still following \cite{Donos:2014uba}, we derive the DC conductivity:
\begin{eqnarray}\label{38}
\sigma_{DC}=\left(1+\frac{\mathcal{K}k^2}{r_h^2}\right)+\frac{\rho^2 }{k^2r_h^{2}\bar{U}(r_h)}.
\end{eqnarray}
This formula agrees with (3.1) of \cite{Baggioli:2016oqk} upon replacing their $Y(X)=1+\mathcal K Tr[X]$.

\begin{figure}
\begin{tabular}{ccc}
\includegraphics[width=0.32\textwidth, trim=1cm 0cm 1cm 0cm]{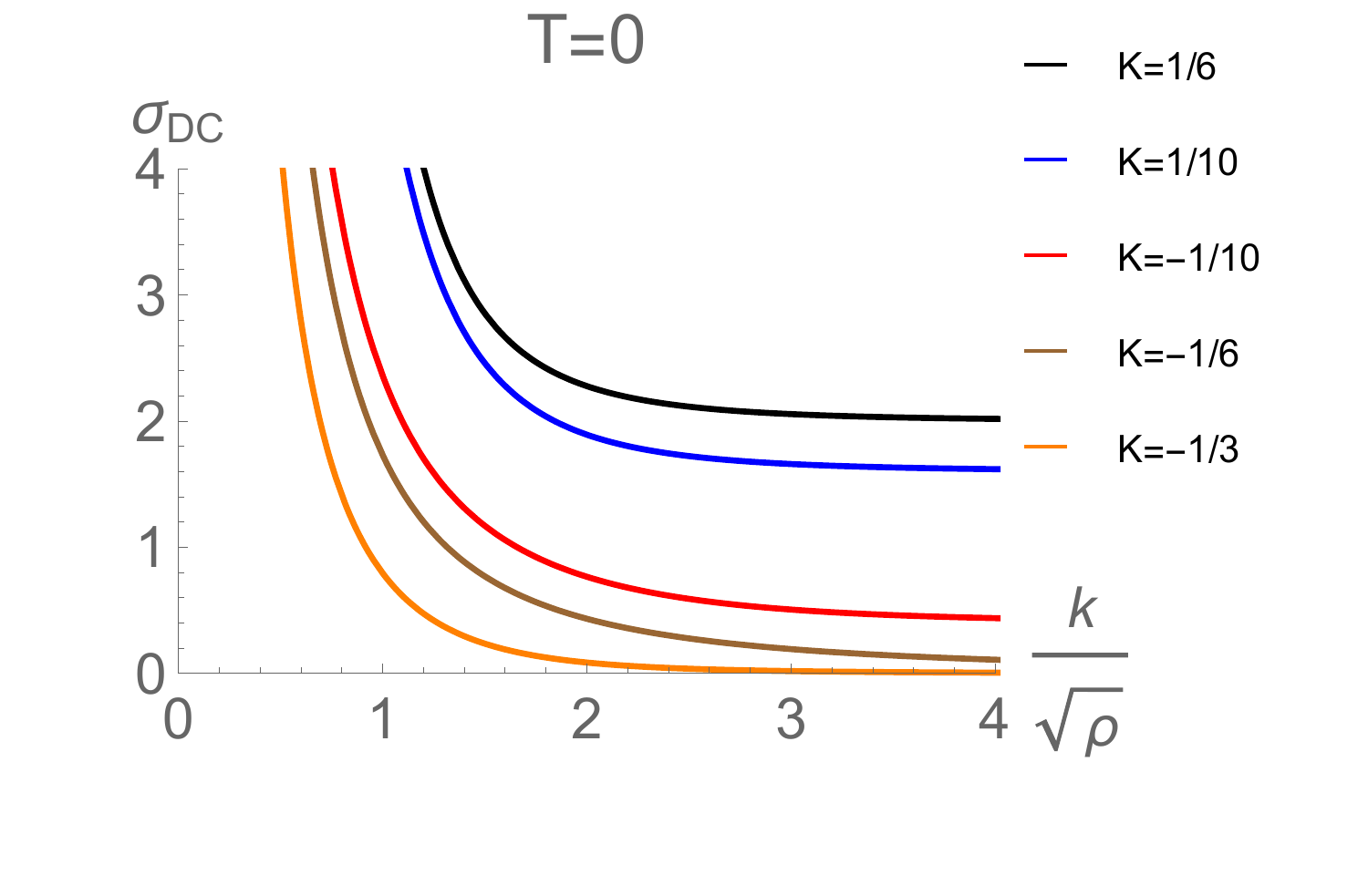} &
\includegraphics[width=0.32\textwidth, trim=1cm 0cm 1cm 0cm]{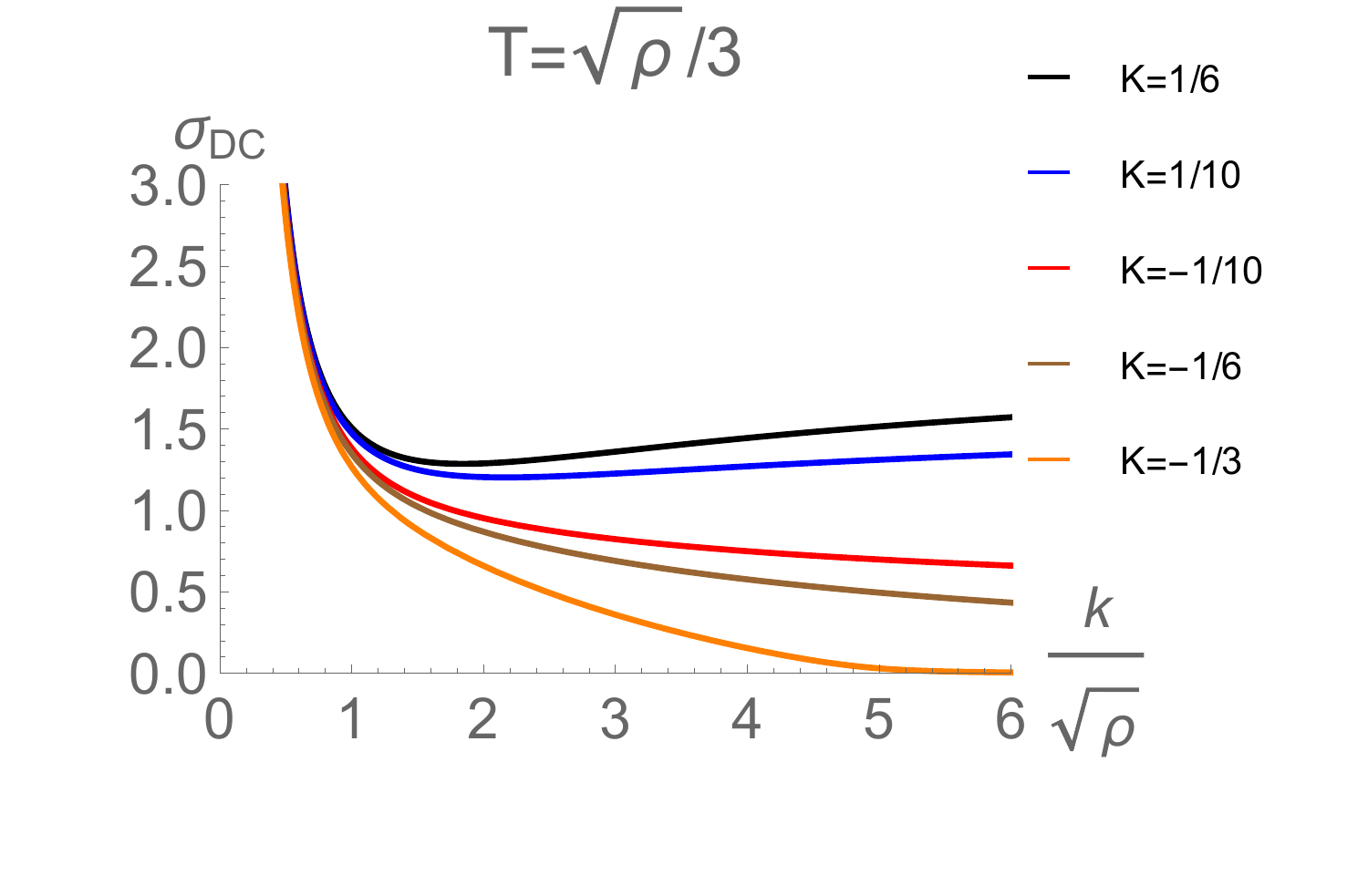}&
 \includegraphics[width=0.32\textwidth, trim=1cm 0cm 1cm 0cm]{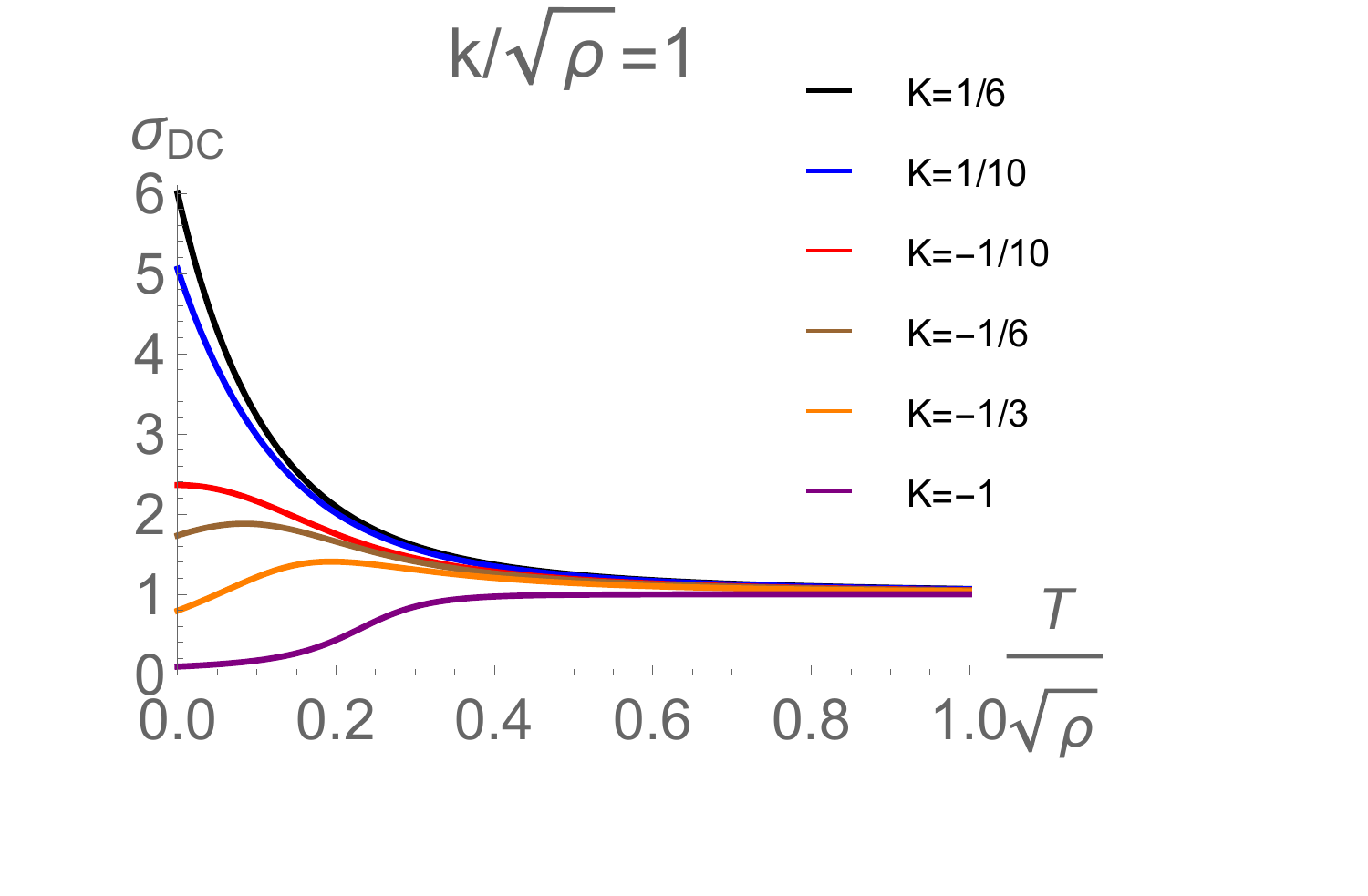}
\end{tabular}
\caption{Plots of the DC conductivity with the $\mathcal K$ coupling turned on, for both positive and negative values. Left, we fix $T=0$ and vary $k/\sqrt\rho$. Center, we fix $T=\sqrt{\rho}/3$ and vary $k/\sqrt\rho$. Right, we fix $k=\sqrt{\rho}$ and vary $T/\sqrt{\rho}$.}
\label{fig:sigmaDC3}
\end{figure}

We plot the DC conductivity in figure \ref{fig:sigmaDC3}.
At zero temperature (keeping $\rho\neq0$, so the black hole becomes extremal), it takes a simple expression
\begin{equation}
\sigma_{DC}(T=0,\tilde k)=\frac{\sqrt{12+\tilde k^4(1+6\mathcal K)^2}}{\tilde k^2(1-6\mathcal K)+\tilde k^4\mathcal K(1+6\mathcal K)\sqrt{12+\tilde k^4(1+6\mathcal K)^2}-\tilde k^6\mathcal K(1+6\mathcal K)^2}
\end{equation}
where $\tilde k=k/\sqrt\rho$.
For $\mathcal K\leq1/6$, the DC conductivity is always positive and monotonously decreasing in terms of $\tilde k$. We note that 
\begin{equation}
\begin{split}
&\sigma_{DC}(T=0,\tilde k)\underset{\tilde k\to+\infty}{\longrightarrow}1+6\mathcal K+O\left(\frac1{\tilde k^4}\right)\quad\textrm{for $\mathcal K>-\frac16$}\,,\\
&\sigma_{DC}(T=0,\tilde k)=\frac{\sqrt{3}}{\tilde k^2}\quad\textrm{for $\mathcal K=-\frac16$}\,,\\
& \sigma_{DC}(T=0,\tilde k)\underset{\tilde k\to+\infty}{\longrightarrow}O\left(\frac1{\tilde k^4}\right)\quad\textrm{for $\mathcal K<-\frac16$}\,.
\end{split}
\end{equation}
For $\mathcal K>1/6$ however, one may check explicitly that the DC conductivity is negative at small $\tilde k$, so we restrict the range of validity of $\mathcal K$ accordingly.

At non-zero temperatures, keeping $\tilde T=T/\sqrt \rho$ fixed and varying $\tilde k$, the conductivity is qualitatively the same as for $T=0$ when $\mathcal K<0$. For $0<\mathcal K\leq1/6$, it is no longer monotonously decreasing: there is a minimum at a finite value of $\tilde k=\tilde k_{min}$.

Keeping $\tilde k$ fixed and varying $\tilde T$, we see an analogous qualitative change of behaviour upon changing the sign of $\mathcal K$. For $\mathcal K>0$, the conductivity is monotonously decreasing. For $\mathcal K$ not too negative, there is a maximum at small temperatures, which disappears as $\mathcal K$ becomes smaller. In both cases, the conductivity at high temperatures asymptotes to $1$, its value at zero density and no translation symmetry breaking.

At finite density, the admissible range of values for $\mathcal K$ then seems to be $\mathcal K\leq1/6$. However, the zero density analysis of appendix \ref{app:schroedcase3} reveals that an infinite negative well forms in the Schr\"odinger potential (signaling an instability) for $\mathcal K<-1/6$. We thus restrict to 
\begin{equation}
-\frac16<\mathcal K\leq \frac16\,.
\end{equation}

\section{Discussion and outlook \label{section4}}

In this note, we have expanded on previous modelisations of holographic homogeneous translation symmetry breaking and charge transport by going beyond the leading low-energy two-derivative terms in the action. This is consistent with the spirit of effective field theories where additional higher-derivative couplings are expected on general grounds provided they are allowed by symmetry. As our emphasis is on charge transport, we have mainly considered higher derivative couplings between the charge and translation symmetry breaking sectors, of the form $\mathcal J Tr[XF^2]$ \eqref{15} and $\mathcal K Tr[X]F^2$ \eqref{28}.

Remarkably, from a purely gravitational point of view, it is still possible to obtain exact analytical charged black hole solutions, going beyond previous literature \cite{Bardoux:2012aw,Andrade:2013gsa,Baggioli:2014roa}. Using a now well-known formalism \cite{Donos:2014uba}, the DC conductivity of the dual field theory is easily derived in terms of horizon data, see formul\ae\ \eqref{26} and \eqref{38} respectively.

Requiring the DC conductivity to be positive at zero temperature for all values of the translation symmetry breaking parameter $k$ allows us to place some constraints on the couplings: $0\leq\mathcal J\leq 2/3$ and $\mathcal K\leq1/6$. Moreover, at zero density, the analysis of the Schr\"odinger problem for the charge fluctuation equation further restricts $\mathcal K>-1/6$ and confirms the upper bound $\mathcal J\leq 2/3$, as explained in appendices \ref{app:schroedcase2} and \ref{app:schroedcase3}. Working out how non zero density affects these constraints is important, but we leave this for future work (it requires decoupling the gauge invariant fluctuations at non-zero density, which is not a trivial task).

In a more general context there are several related issues that need to be clarified
\begin{enumerate}

\item Physicality constraints on  the effective field theories. These involve unitarity constrains as well as superluminality constraints \cite{Brigante:2007nu,Brigante:2008gz,Buchel:2009sk}. Some of them were applied in this work, but it is important to know the full set of such constraints and how to obtain them. The approach of \cite{Adams:2006sv} may be useful in this respect.

\item In many cases we use the $k\to\infty$ limit of such solutions as the limit of strong relaxation. However in string theory this limit does not exist as such for several reasons. The main ingredient is that scalars without a potential have very peculiar properties in string theory. Exactly marginal operators like the Type IIB axion have indeed no potential. This is the text book example of an exactly-marginal operator. However, both at the two-derivative level and most importantly at the higher derivative level their couplings are modular forms of the axion-dilaton, and therefore for solutions that source their kinetic terms as the ones described here, the dilaton is sourced
    and there is a non-trivial flow. Therefore such solutions will need to be modified.

 In other less symmetric theories, the axions obtain a potential from instantons that may be not exponentially small in $N$. In such a case the linear axion solutions are approximate.

 Finally, for fields with global symmetries like the Q-lattices \cite{Donos:2013eha}, it is well known that such symmetries are not exact, as there are no global symmetries in string theory. There can be typically small violations of such symmetries that may be due to stringy or non-perturbative effects and therefore even in such cases such solutions are approximate.
It would be interesting to establish model independent bounds on $k$ where such simple solutions can be trusted.
Till then effects happening at the $k\to\infty$ limit may not be trustworthy.

\item There is finally the possibility that in describing momentum relaxation using axions, a certain coarse-graining procedure is involved. Indeed, depending on dimension such solutions can be interpreted as a continuum limit of discrete brane distributions. The presence of a brane at codimension one shifts the value of the dual axion field in the perpedicular direction \cite{Kiritsis:2007zza}.
In the presence of a regular array of such branes and in the continuum limit, the axion solutions relevant here are obtained. We cannot exclude therefore some loss of unitarity in this limit, which may lead to relax some of the constraints on the conductivity. A physical interpretation would be needed as guidance in such a case.

\end{enumerate}

We now come back to the lower bound on the conductivity proven by \cite{Grozdanov:2015qia}. It applies to four-dimensional Einstein-Maxwell-AdS theories coupled to scalars with potentials, which in particular  contains the model of \cite{Andrade:2013gsa}. \cite{Grozdanov:2015qia} found that the conductivity was bounded from below $\sigma_{DC}\gtrsim \sigma_0$ for arbitrary strength of translation symmetry breaking, and that the bound corresponded to the conductivity of the clean, neutral CFT plasma $\sigma_0=1$. Our results show that upon including higher derivative couplings between the charge and translation symmetry breaking sectors, no lower bound survives in general. Indeed, in both cases the DC conductivity can get arbitrarily close to zero at large translation symmetry breaking $k$ and zero temperature. This happens either when $\mathcal J=2/3$ or $\mathcal K\to-1/6$. Away from these values and for the $\mathcal J$ coupling, the DC conductivity at zero temperature is bounded from below by its asymptotic value at infinite strength of translation symmetry breaking. In the case of the $\mathcal K$ coupling, the DC conductivity has a more complicated behaviour, is not always monotonous, and is not always bounded from below by its asymptotic value for infinite strength of translation symmetry breaking. These asymptotic values are always different from the conductivity of the neutral, clean CFT plasma, which is still $\sigma_0=1$ in our setup.

Our results also mean the phases described in this paper should be thought of as metallic, as the DC conductivity is never zero at zero (or non-zero) temperature except asymptotically for specific values of the coupling. So while there is no lower bound, the issue of whether a metal to insulator phase transition (e.g. the conductivity vanishes) can happen at fixed temperature in terms of the strength of translation symmetry breaking without violating some stability constraint remains open.

There are several future directions. It would be very worthwhile to consider couplings between gravity and the translation symmetry breaking sector, which should now modify the heat conductivity compared to the results of \cite{Donos:2014cya}. Another bound was proposed in \cite{Grozdanov:2015djs} and it might also be sensitive to these higher derivative couplings.

In \cite{Davison:2015bea}, it was shown that without higher derivative couplings, there was a map between the bulk gauge invariant decoupled modes and the coherent/incoherent currents in the boundary theory. Our solutions are simple enough that it may be possible to work out the decoupling of modes, and study the effect of the higher derivative couplings on the momentum relaxation rate for instance. More generally, the frequency behaviour of the conductivity is interesting, in particular in view of particle-vortex duality (which would amount to flipping the sign of $\mathcal K$ for small enough values of $|\mathcal K|$), \cite{Herzog:2007ij,Myers:2010pk,WitczakKrempa:2012gn,WitczakKrempa:2013ht,Witczak-Krempa:2013nua,Witczak-Krempa:2013aea}.

\acknowledgments

The authors would like to thank Matteo Baggioli, Oriol Pujolas and Andy Lucas for comments on the first version of the manuscript.

 B.G. would like to thank CCTP and QCN  at the University of Crete for warm hospitality over the course of this work.

This work was supported in part by European Union's Seventh Framework Programme under grant agreements (FP7-REGPOT-2012-2013-1) no 316165, and the Advanced ERC grant SM-grav, No 669288. The work of B.G. is supported by the Marie Curie International Outgoing Fellowship nr 624054 within the 7th European Community Framework Programme FP7/2007-2013.

\appendix
\section{Covariant form of the equations of motion\label{section5}}
We start with the general effective holographic action (\ref{10})-(\ref{12}) and set $d=2$. Varying $A_{\mu}$, $X^I$ and $g_{\mu\nu}$, the equations of motion are given by
\begin{eqnarray}\label{a1}
\nabla_\mu\sum_{n,m=0}^{\infty}Z_{n,m}\Big[Tr[X^m]\Big((X^nF)^{\mu\nu}-(X^nF)^{\nu\mu}+(FX^n)^{\mu\nu}-(FX^n)^{\nu\mu}\Big)\Big]=0\,,
\end{eqnarray}
\begin{eqnarray}\label{a2}
&&\nabla_\mu \Big\{ \sum_{l=1}^{\infty}l(f_l R-V_{X^l})\left({(X^{l-1})^\mu}_\nu\nabla^\nu X^I+\nabla_vX^I(X^{l-1})^{\nu\mu}\right)\\ \nonumber
&&-\frac{1}{4}\Big[\sum_{n=0,m=0}^{\infty}mZ_{n,m}\Big({(X^{m-1})^\mu}_\nu\nabla^\nu X^I
+\nabla_vX^I(X^{m-1})^{\nu\mu}\Big)Tr[X^nF^2]\\ \nonumber
&&+Z_{n,m}Tr[X^m]\sum_{a=0}^{n-1}\left({(X^{n-a-1}F^2X^a)^\mu}_\nu\nabla^\nu X^I+\nabla_\nu X^I (X^{n-a-1}F^2X^a)^{\nu\mu}\right)\Big]\Big\}=0\,,
\end{eqnarray}
and
\begin{eqnarray}\label{a3}
&&\Big(1+\sum_{l=1}^{\infty}f_l Tr[X^l] \Big)(R_{\mu\nu}-\frac{1}{2}g_{\mu\nu}R)+\sum_{l=1}^{\infty}(f_l R-V_{X^l})G^{(l)}_{\mu\nu}(X^I,g)-\frac{1}{2}(6-V(X))g_{\mu\nu}\nonumber\\
&&=\frac{1}{4}\sum_{n,m=0}^{\infty}Z_{n,m}\Big[G_{\mu\nu}^{(m)}(X,g)Tr[X^nF^2]+Tr[X^m] H^{(n)}_{\mu\nu}(X^I,g)-\frac{1}{2}g_{\mu\nu}Tr[X^m]Tr[X^nF^2]\Big]\,,
\end{eqnarray}
where
 $\nabla_\mu$ is the covariant derivative, 
 \begin{equation}
G_{\mu\nu}^{(0)}(X^I,g)=0\,,\qquad G_{\mu\nu}^{(n>0)}(X^I,g)\equiv\frac{n}{2} \nabla_{(\mu|} X^I\nabla_\alpha X^I{(X^{n-1})^{\alpha}}_{|\nu)}\,,
 \end{equation} 
 \begin{equation}
 H_{\mu\nu}^{(0)}(X^I,g)=2{F_\mu}^\lambda F_{\lambda \nu}
 \end{equation}
  and 
 \begin{equation}
 H_{\mu\nu}^{(n>0)}(X,g)\equiv \frac{1}{2} \nabla_{(\mu|} X^I\nabla_\alpha X^I\sum^{n-1}_{a=0}{(X^{n-a-1}F^2X^a)^\alpha}_{|\nu)}\\+F_{(\mu|\lambda}{(FX^n)^\lambda} _{|\nu)}+{F_{(\mu|\lambda}(X^nF)^\lambda} _{|\nu)}\,.
 \end{equation}
 Here we have adopted the notation $A_{(\mu|...|\nu)}\equiv\frac{1}{2}\left(A_{\mu...\nu}+A_{\nu...\mu}\right)$.

For the special case $V(X)=Tr[X]$, $Z_{0,0}=-1$, $Z_{0,1}=-\mathcal{K}$, $Z_{1,0}=\mathcal{J}$ without other couplings, the equations above reduce to
\begin{eqnarray}\label{a4}
\nabla_\mu\left[\left(1+\frac{\mathcal{K}}{2}\nabla_\sigma X^I \nabla^\sigma X^I\right)F^{\mu\nu}-\frac{\mathcal{J}}{2}\left((XF)^{\mu\nu}-(XF)^{\nu\mu}\right)\right]=0\,,
\end{eqnarray}
\begin{eqnarray}\label{a5}
&&\nabla_\mu \Big[\nabla^\mu X^I-\frac{1}{4}\left(\mathcal{K}\nabla^\mu X^I Tr[F^2]-\mathcal{J}{(F^2)^\mu}_\nu\nabla^\nu X^I\right)\Big]=0\,,
\end{eqnarray}

and

\begin{eqnarray}\label{a6}
&&R_{\mu\nu}-\frac{1}{2}g_{\mu\nu}R-\frac{1}{2}(1+\frac{\mathcal{K}}{4}F_{\sigma\rho}F^{\sigma\rho})\nabla_\mu X^I\nabla_\nu X^I-\frac{1}{2}\left(6-\frac12\nabla_\sigma X^I \nabla^\sigma X^I\right)g_{\mu\nu}\\ \nonumber
&&=\frac{1}{2}\left(1+\frac{\mathcal{K}}{2}\nabla_\sigma X^I \nabla^\sigma X^I\right)\Big({F_\mu}^{\sigma}F_{\nu\sigma}-\frac{1}{4}g_{\mu\nu}F_{\rho\sigma}F^{\rho\sigma}\Big)+\frac{\mathcal{J}}{4}\Big(\frac{1}{2}\nabla_{(\mu|} X^I\nabla_\sigma X^I{(F^2)^{\sigma}}_{|\nu)}\\ \nonumber
&&+F_{(\mu|\sigma}{(FX)^{\sigma}}_{|\nu)}+F_{(\mu|\sigma}{(XF)^{\sigma}}_{|\nu)}-\frac{1}{2}g_{\mu\nu} Tr[XF^2]\Big).
\end{eqnarray}

\section{Schr\"odinger analysis for the $Tr[X F^2]$ coupling at zero density \label{app:schroedcase2}}

At zero density $\rho=0$, the $\mathcal J$ coupling does not enter in the Einstein or scalar equations, only in the $x$ component of the linearized Maxwell equations. The $a_x$ perturbation also decouples from the other parity-odd perturbations and obeys the following equation at non zero frequency $\omega$ and zero wavevector $p=0$:
\begin{equation}
\label{eq:axeomJrho=0}
\left[\left(1-\frac{\mathcal{J}k^2}{4r^2}\right)D a_x'\right]'+\left(1-\frac{\mathcal{J}k^2}{4r^2}\right)\frac{\omega^2 }{D}a_x=0\,.
\end{equation}

\begin{figure}
\begin{tabular}{cc}
\includegraphics[width=0.47\textwidth]{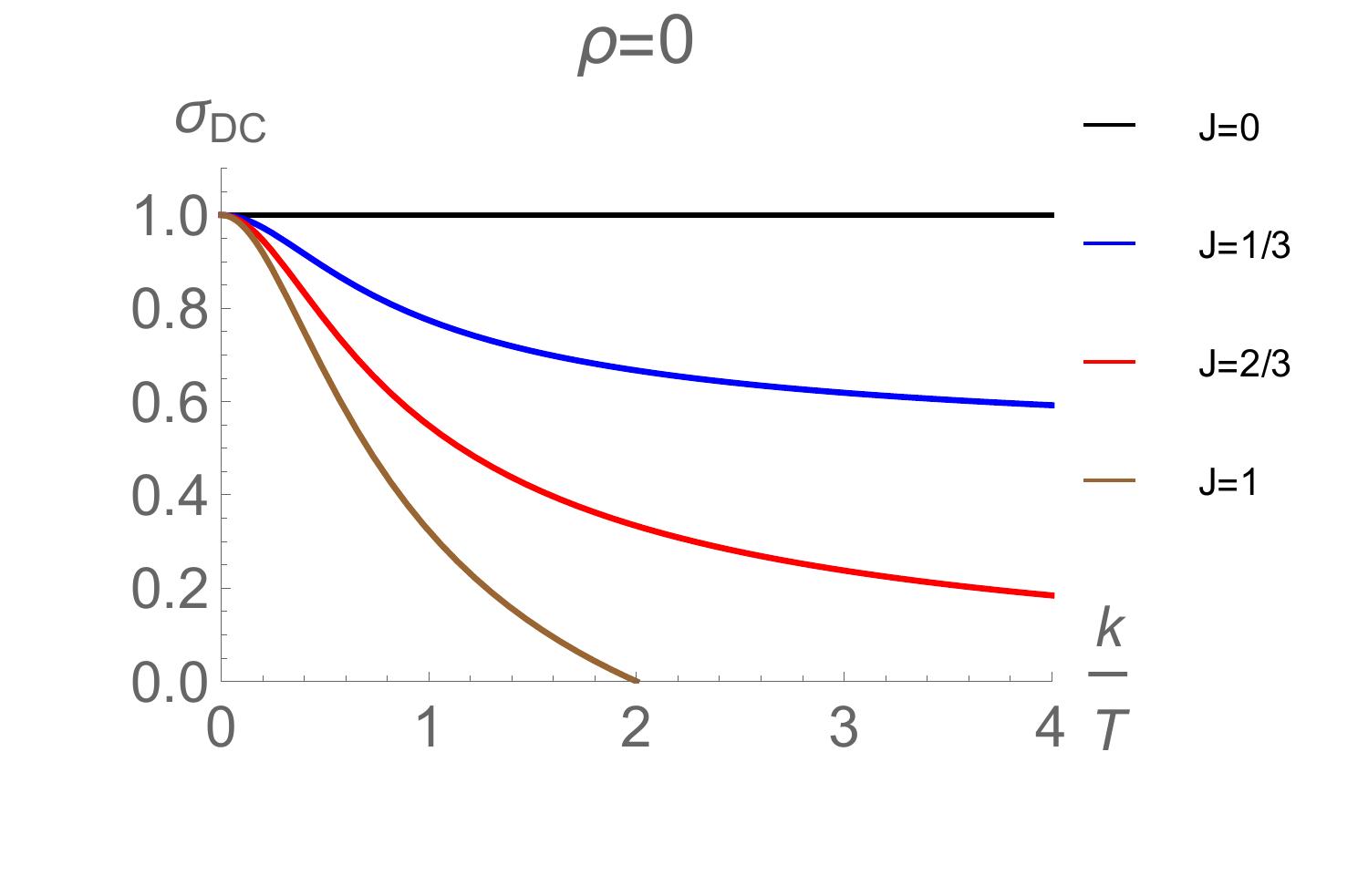}&
\includegraphics[width=0.47\textwidth]{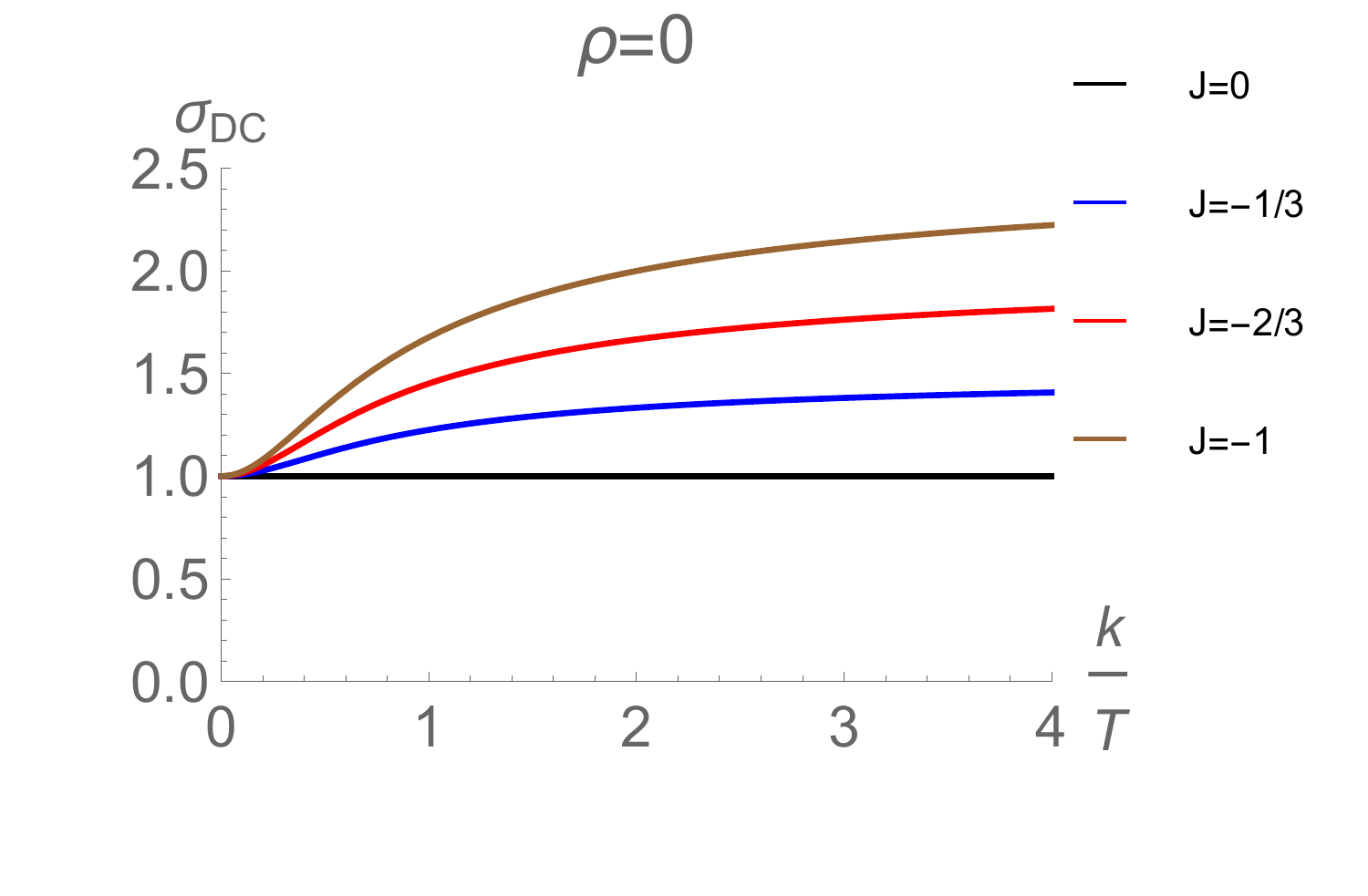}
\end{tabular}
\caption{Plot of the DC conductivity \eqref{sigmaDC2rho0} with the $\mathcal J$ coupling turned on at zero density $\rho=0$, for positive (left) and negative (right) values of $\mathcal J$.}
\label{fig:sigmaDC2vskrho=0}
\end{figure}

Moreover, in this limit the DC conductivity is
\begin{equation}
\label{sigmaDC2rho0}
\sigma_{DC}^{\rho=0}=1-\frac{\mathcal{J}k^2}{4 r_h^2}\underset{\tilde k\to+\infty}{\longrightarrow}1-\frac{3\mathcal J}{2}
\end{equation}
and as seen in figure \ref{fig:sigmaDC2vskrho=0}, it vanishes at finite $\tilde k=k/4\pi T$ for $\mathcal J>2/3$. Unitarity of the dual field theory requires the conductivity to be positive, which is consistent with the range $\mathcal J\leq2/3$.\footnote{Contrarily to the finite density analysis, $\mathcal J<0$ is not disqualified as $\sigma_{DC}(T=0)>0$ in this case.} As at non-zero density, when the inequality is saturated the conductivity vanishes asymptotically.

At zero density, charge transport in Einstein-Maxwell-Ads theories in $d=2$ can be understood in  terms of a particle-vortex duality, related to the self-duality of the Maxwell term in the action \cite{Herzog:2007ij}. In the presence of higher-derivative couplings between the gravity and charge sector, this notion is formally lost, but can be restored by a flip of the sign of the coupling \cite{Myers:2010pk}. This extended notion of particle-vortex duality is still very useful to understand charge dynamics in the higher-derivative system \cite{WitczakKrempa:2012gn,WitczakKrempa:2013ht,Witczak-Krempa:2013nua,Witczak-Krempa:2013aea}. It is plausible that this would also apply to our setup, for values of $-2/3<\mathcal J<2/3$, given the apparent symmetry of the DC conductivity within this range.

To determine whether there are further constraints on the value of $\mathcal J$, it is useful to change variables to
\begin{equation}
\frac{dz}{dr}=\frac{1}{D(r)}\,,\quad a_x(r)=\frac{\Psi(r)}{\sqrt{1-\frac{\mathcal J k^2}{4 r^2}}}
\end{equation}
so as to recast equation \eqref{eq:axeomJrho=0} as a Schr\"odinger equation
\begin{equation}
-\partial_z^2 \Psi(z)+V_{schr}(z)\Psi(z)=\mathcal E \Psi(z)\,,\quad \mathcal E=\omega^2\,.
 \end{equation}
The rationale behind this is to use our intuition from solutions to the Schr\"odinger equation to determine whether (metastable) bound states can form whenever there is a deep enough negative well in the Schr\"odinger potential. The formation of a bound state signals an instability of the neutral CFT plasma and is accompanied by the crossover of a pole from the lower half complex frequency plane to the upper half plane.

This may be understood as follows \cite{Myers:2007we}. Given that we are imposing ingoing boundary conditions at the horizon (which can be checked is at $z\to+\infty$), the wavefunction there will obey $\Psi(z)\sim \exp(i\omega z)$ and vanish at the boundary. Typically this means that the eingenvalues for the effective energy $\mathcal E=\omega^2$ will be complex. This matches the intuition that the quasinormal modes of the black hole are also complex with $\omega=\pm\Omega-i\Gamma$. A stable mode will have $\Gamma>0$, so that it decays in time close to the horizon.

However, if a bound state forms at the bottom of a negative well of the Schr\"odinger potential, then the wave function will see a potential barrier before reaching the horizon and will exponentially decay for solutions regular at the horizon, $\Psi\sim \exp(-|\Gamma|z)$. Given the ingoing boundary condition $\Psi(z)\sim \exp(i\omega z)\sim\exp(\Gamma z)$, this implies $\Gamma<0$ which contradicts our previous requirement that these modes decay in time and signals an instability. Of course, this is only indicative of what the real spectrum looks like and does not preclude a full quasinormal mode analysis.

Using a WKB approximation, a zero energy bound state will form in the negative potential well for each integer $n$ such that
\begin{equation}
\label{BoundStateFormation}
(n-\frac12)\pi=\int dz \sqrt{-V_{schr}(z)}
\end{equation}
where the integral has support in the region of negative Schr\"odinger potential.

\begin{figure}
\begin{tabular}{cc}
\includegraphics[width=0.47\textwidth]{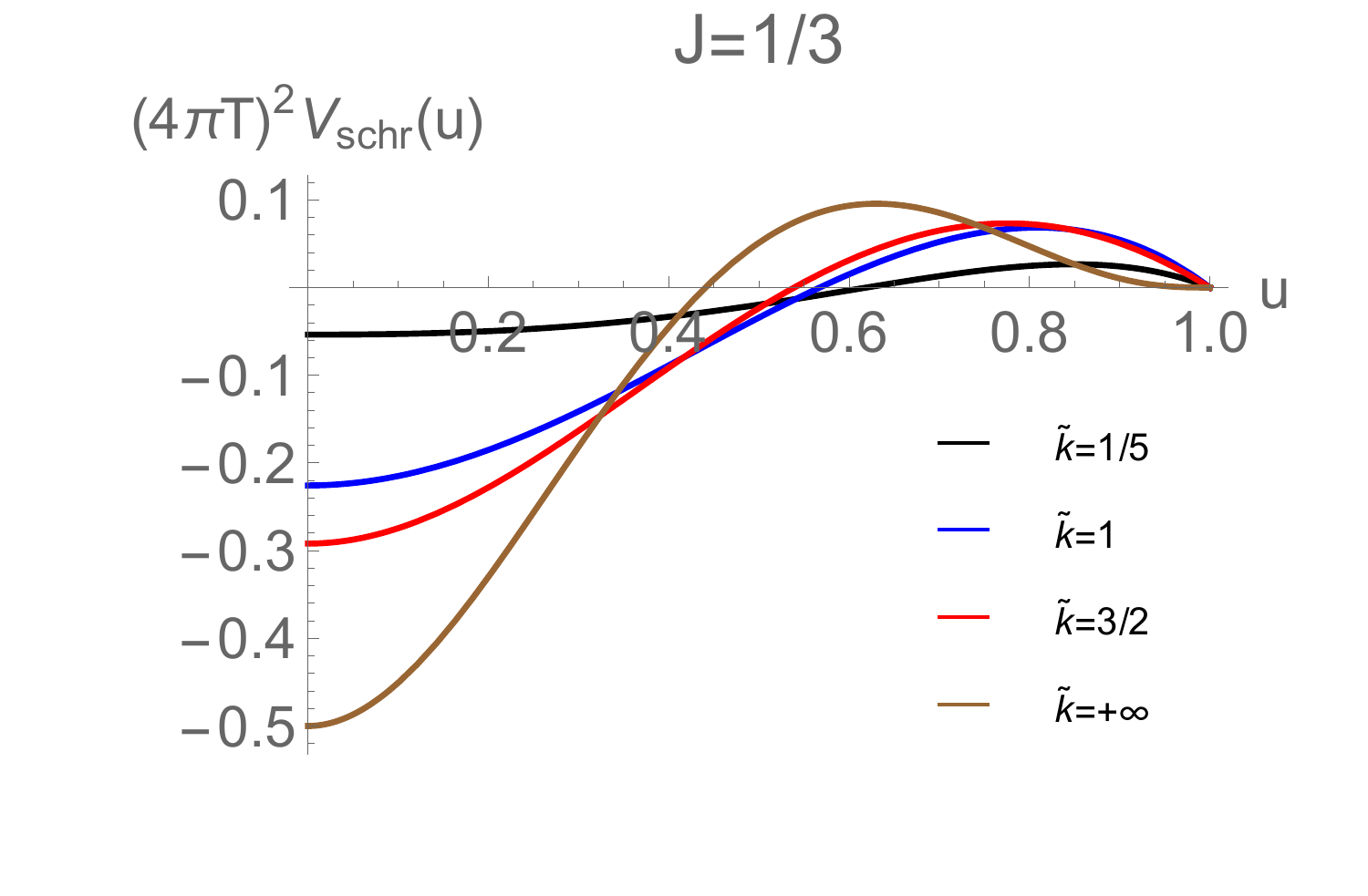}&
\includegraphics[width=0.47\textwidth]{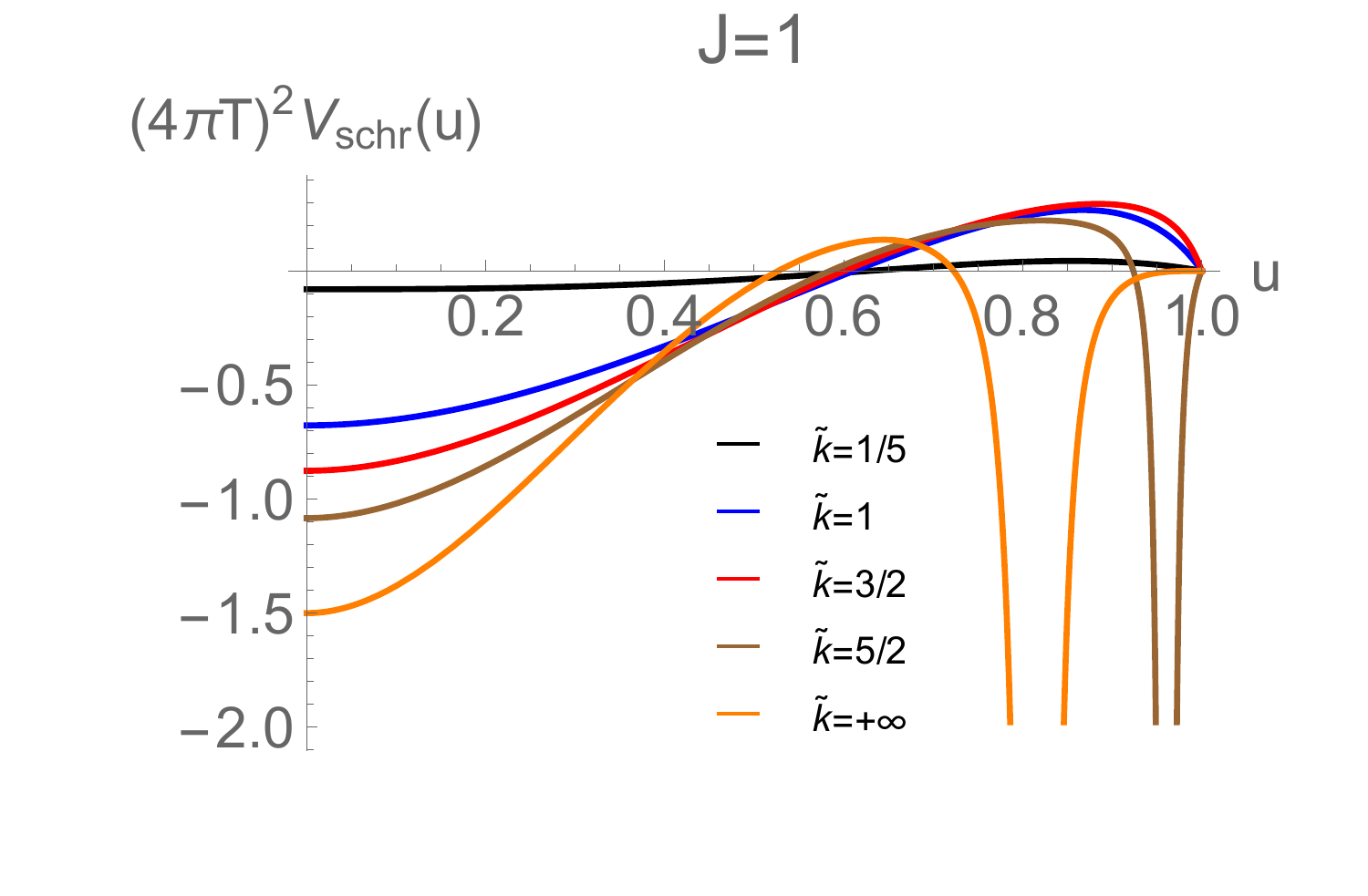}\\
\includegraphics[width=0.47\textwidth]{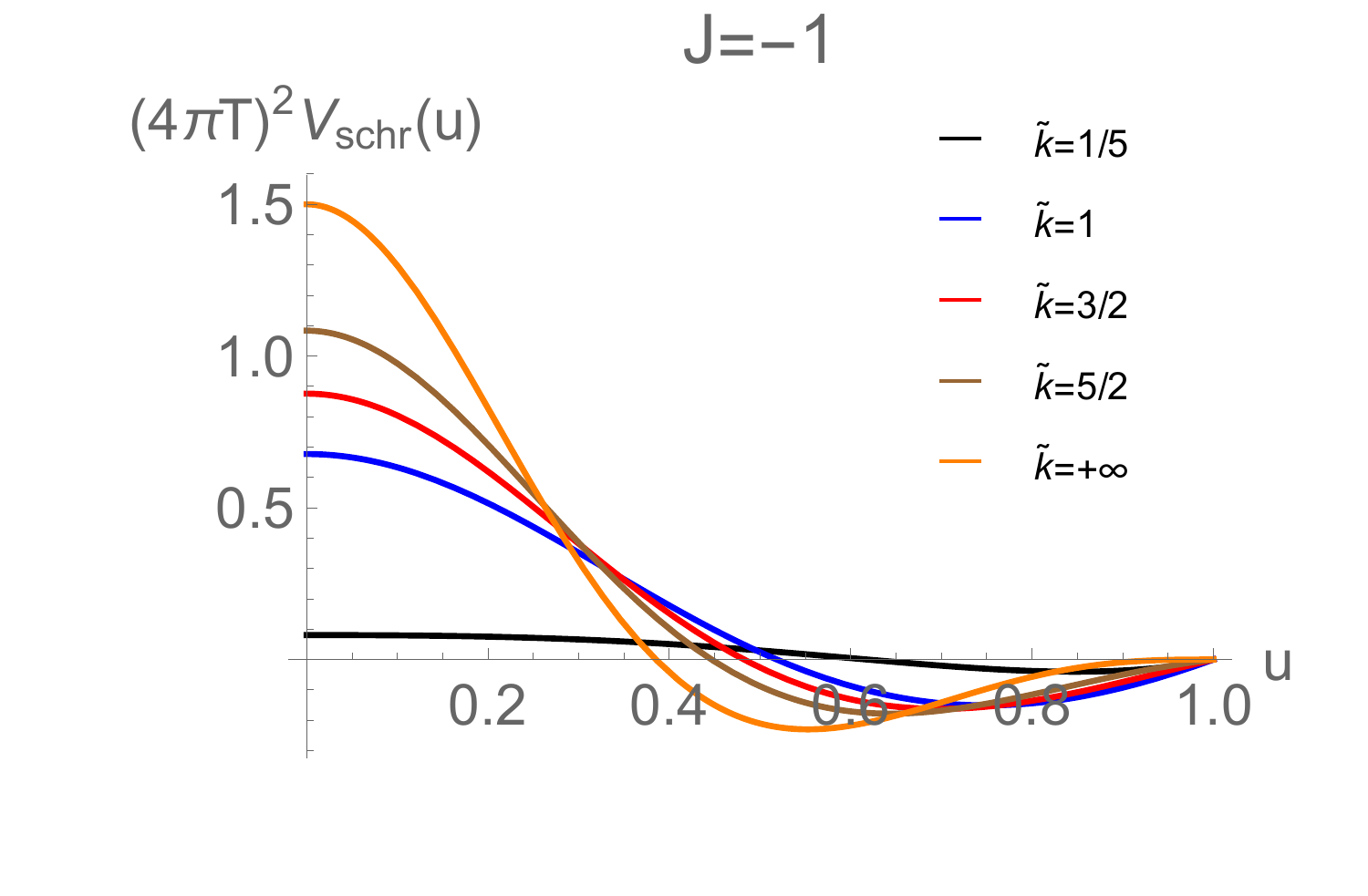}&
\includegraphics[width=0.47\textwidth]{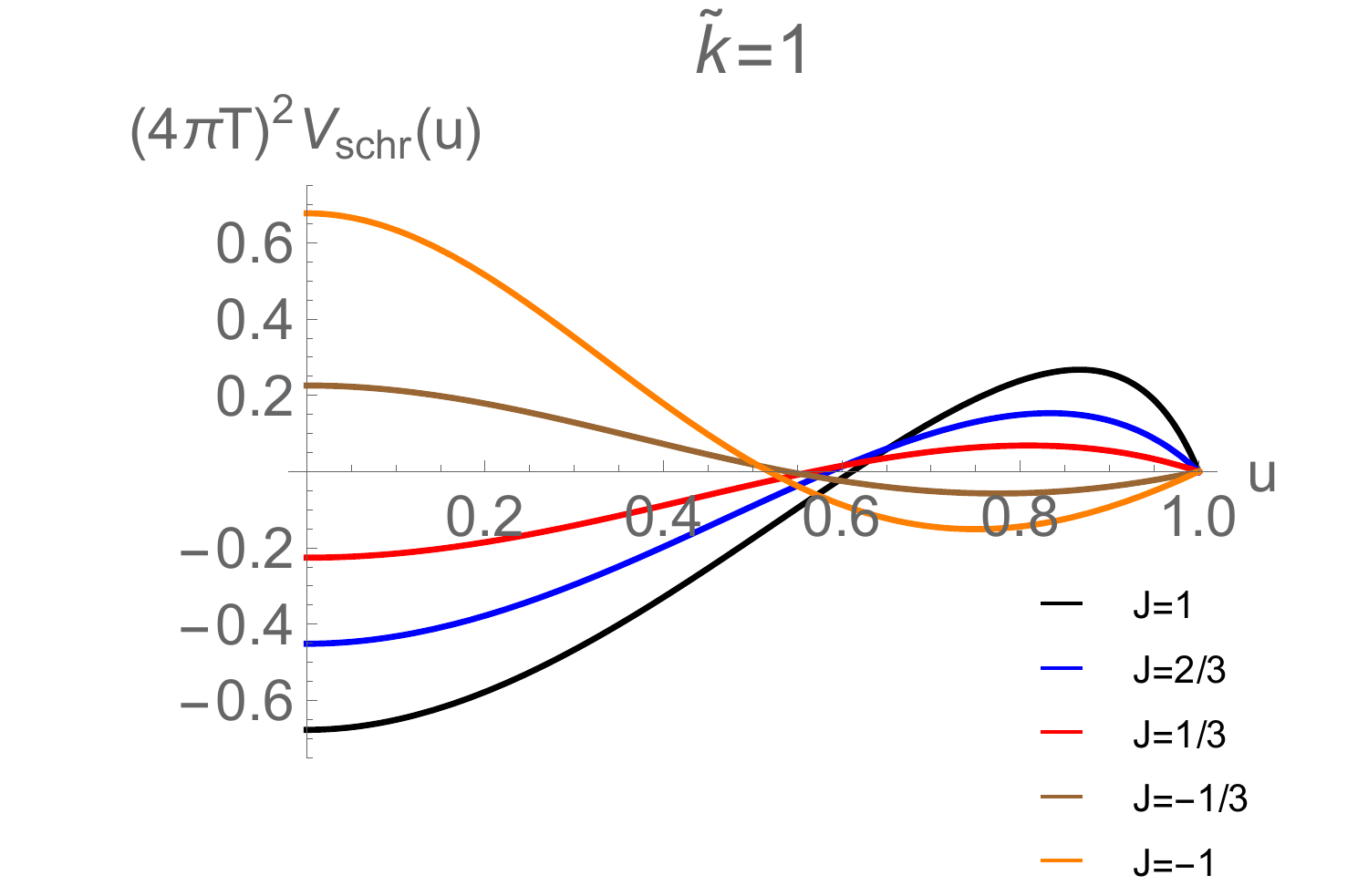}
\end{tabular}
\caption{Plots of the Schr\"odinger potential \eqref{Vschrcase2} versus the radial coordinate $0<u<1$ in units of temperature, with $\tilde k=k/4\pi T$. Top row, for positive values of $\mathcal J$: left for $0<\mathcal J\leq2/3$, showing the presence of a negative well in the potential at the boundary, and a maximum close to the horizon; right, for $\mathcal J>2/3$, showing that for large enough $k$, an infinite negative well develops. Bottom row left, for $\mathcal J=-1$, a positive but finite potential barrier appears at the boundary with a negative well closer to the horizon. Bottom row, right, we fix $\tilde k=1$ and show the dependence of the height/depth of positive/negative potential barrier on the value of $\mathcal J$.}
\label{fig:SchroedPot2vskrho=0}
\end{figure}

In our case, the Schr\"odinger potential reads
\begin{equation}
\label{Vschrcase2}
V_{schr}(r)=\frac{ \mathcal J k ^2D(r) \left[r D'(r) \left(4 r^2-k ^2 \mathcal J\right)+2 D(r) \left(k ^2 \mathcal J-6 r^2\right)\right]}{\left(k ^2 \mathcal J r-4 r^3\right)^2}\,.
\end{equation}
We plot it in figure \ref{fig:SchroedPot2vskrho=0} both for positive and negative $\mathcal J$, using a radial coordinate $u=r_h/r$ (so that the boundary is at $u=0$ and the horizon at $r=r_h$). Note that the potential always vanishes on the horizon, consistent with our boundary condition there. For $0<\mathcal J<2/3$, the Schr\"odinger potential is negative close to the boundary and there is a positive potential barrier before reaching the horizon. As a consequence, a bound state might form there if the well is deep enough.\footnote{Other examples of possible bound state formation and negative wells in the Schr\"odinger potential for the conductivity have been reported in \cite{Bhattacharya:2014dea}.} However, as the well is deepest for $\mathcal J=2/3$ and $\tilde k\to+\infty$, we have explicitly checked that there is no integer such that \eqref{BoundStateFormation} is verified. For $\mathcal J>2/3$, the denominator of the Schr\"odinger potential blows up at $u_\star=2(1+3 \tilde k^2+\sqrt{1+6\tilde k^2})/(9\mathcal J\tilde k^2)$, which is located between $0$ and $1$ for large enough $\tilde k>2\sqrt{\mathcal J}/(3\mathcal J-2)$. Any wave incoming from the boundary will then see an infinite well and a bound state will form, signalling an instability of the neutral CFT plasma for $\mathcal J>2/3$, consistent with the positivity condition of the DC conductivity.

For negative values of $\mathcal J$, we see on the contrary that there is a positive potential barrier at the boundary. Its height becomes larger as $\mathcal J$ is more negative. However, in the spirit of effective field theories, the coupling should remain small. Closer to the horizon, there is a negative well. The depth of this well is maximum at $\tilde k=+\infty$ and also increases with lower values of $\mathcal J$. Using \eqref{BoundStateFormation}, we have estimated that the first zero energy bound state will form for values of $\mathcal J\lesssim -1.188$.

To summarize, the allowed range of $\mathcal J$ at zero density is $-1.188 \lesssim \mathcal J\leq 2/3$.

\section{Schr\"odinger analysis for the $Tr[X] F^2$ coupling at zero density \label{app:schroedcase3}}

\begin{figure}
\begin{center}
\includegraphics[width=0.47\textwidth]{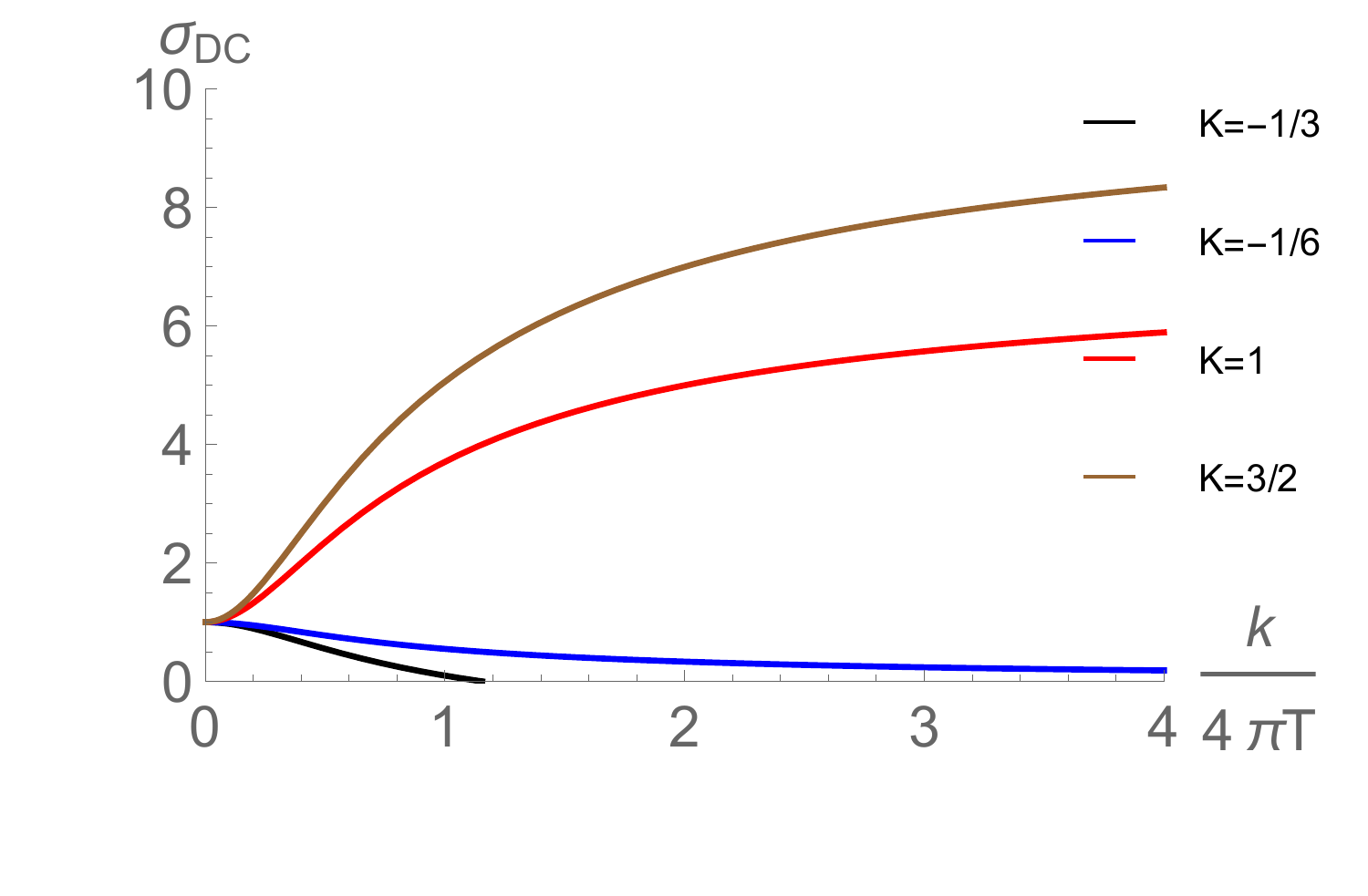}
\end{center}
\caption{Plot of the DC conductivity \eqref{sigmaDC3rho0} with the $\mathcal K$ coupling turned on at zero density $\rho=0$, for positive and negative  values of $\mathcal K$.}
\label{fig:sigmaDC3vskrho=0}
\end{figure}

Let us now repeat the analysis of appendix \ref{app:schroedcase2} applied to the $\mathcal K$ coupling, at zero density. This is facilitated by the fact that the formul\ae\ are the same upon replacing $\mathcal J=-4\mathcal K$. The DC conductivity \eqref{38} simplifies to:
\begin{equation}
\label{sigmaDC3rho0}
\sigma_{DC}^{\rho=0}=1+\frac{\mathcal K k^2}{r^2}
\end{equation}
which we plot as a function of $\tilde k=k/4\pi T$ for representative values in figure \ref{fig:sigmaDC3vskrho=0}. We note that the DC conductivity is positive at zero temperature for all values of $\mathcal K>-1/6$. For $\mathcal K<-1/6$, it vanishes for a finite $k$. The same remarks as in appendix \ref{app:schroedcase2} apply about the role of particle-vortex duality: it is plausible that for $-1/6<\mathcal K<1/6$.

\begin{figure}
\begin{tabular}{cc}
\includegraphics[width=0.47\textwidth]{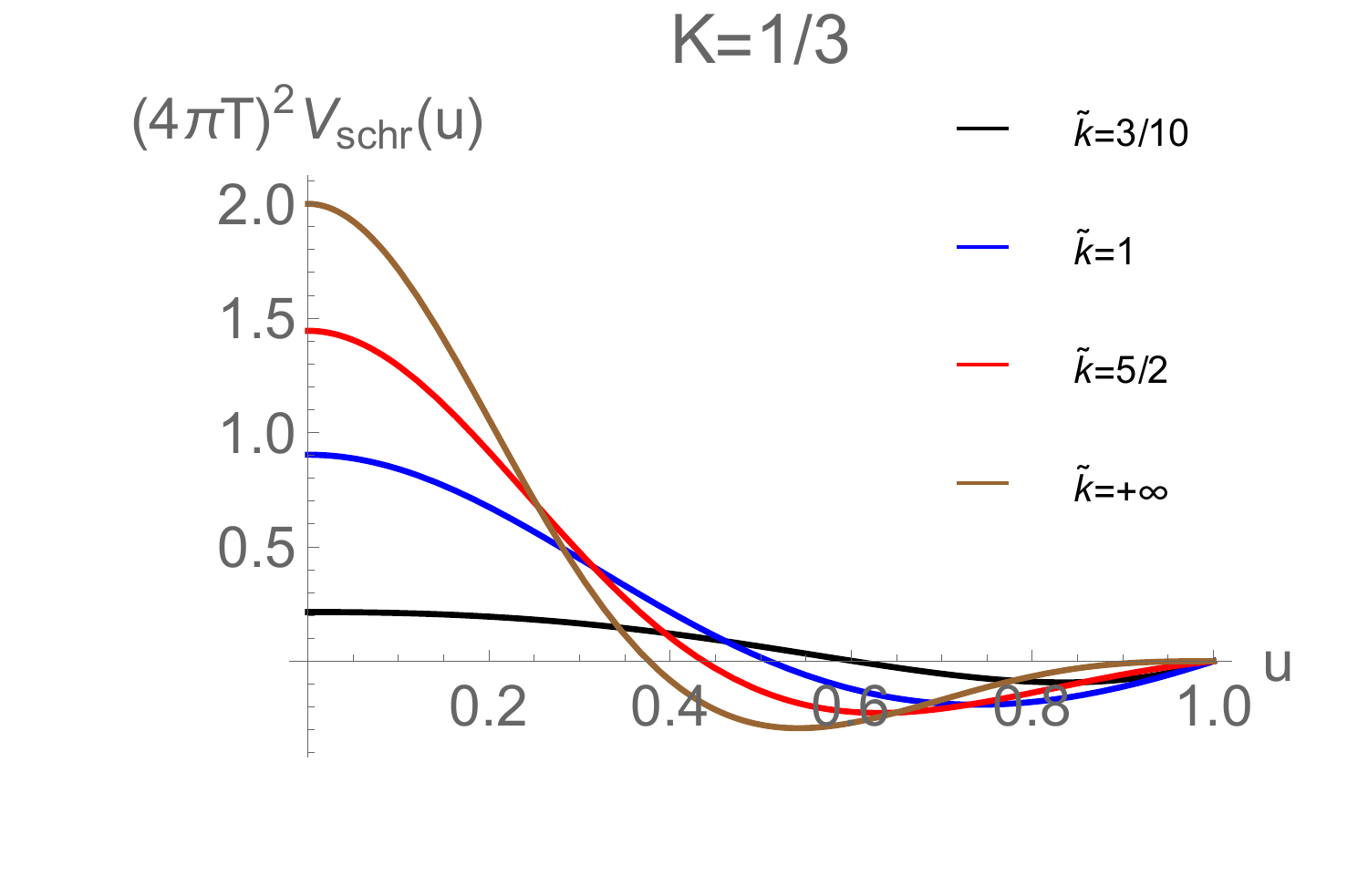}&
\includegraphics[width=0.47\textwidth]{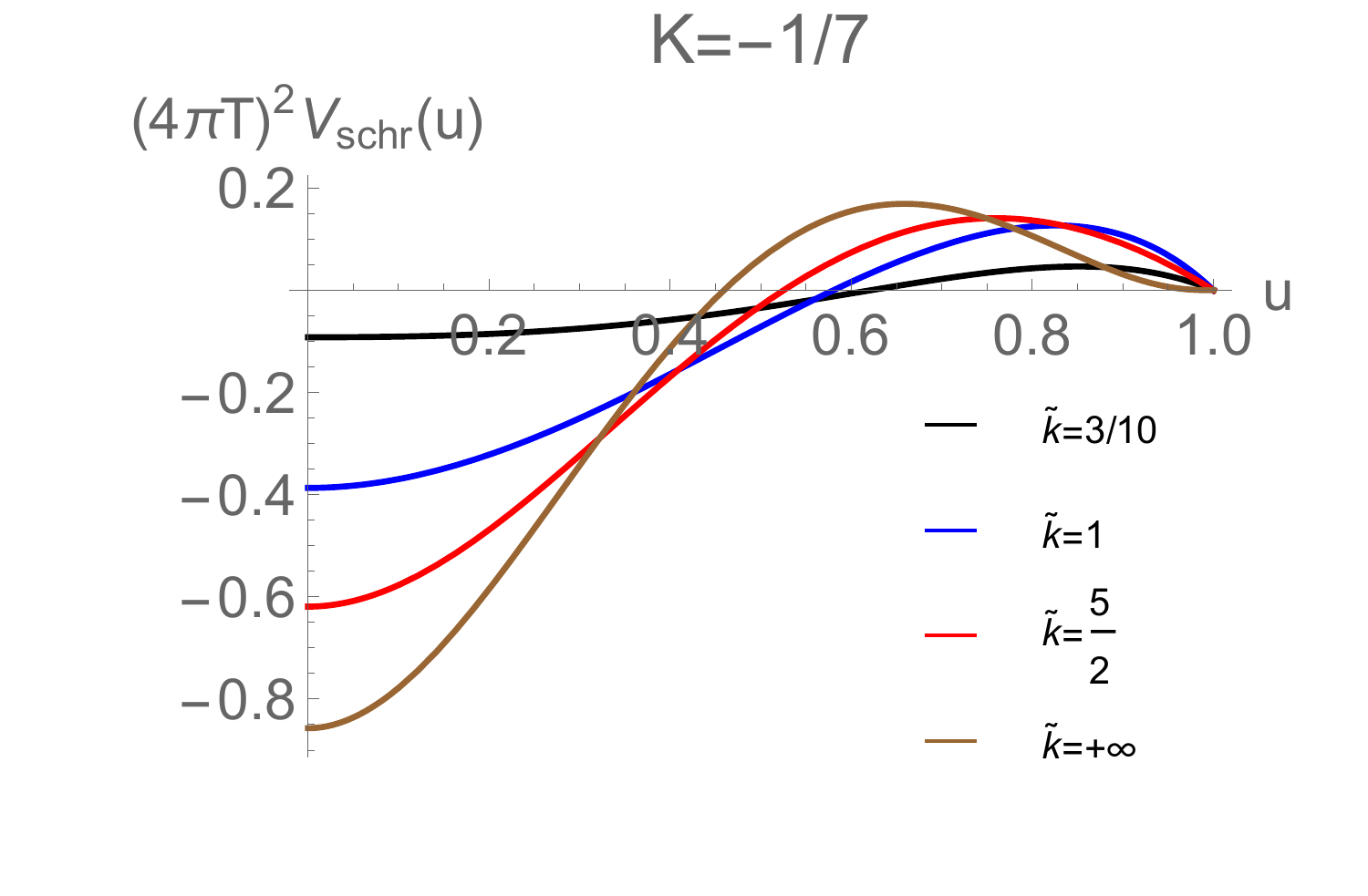}\\
\includegraphics[width=0.47\textwidth]{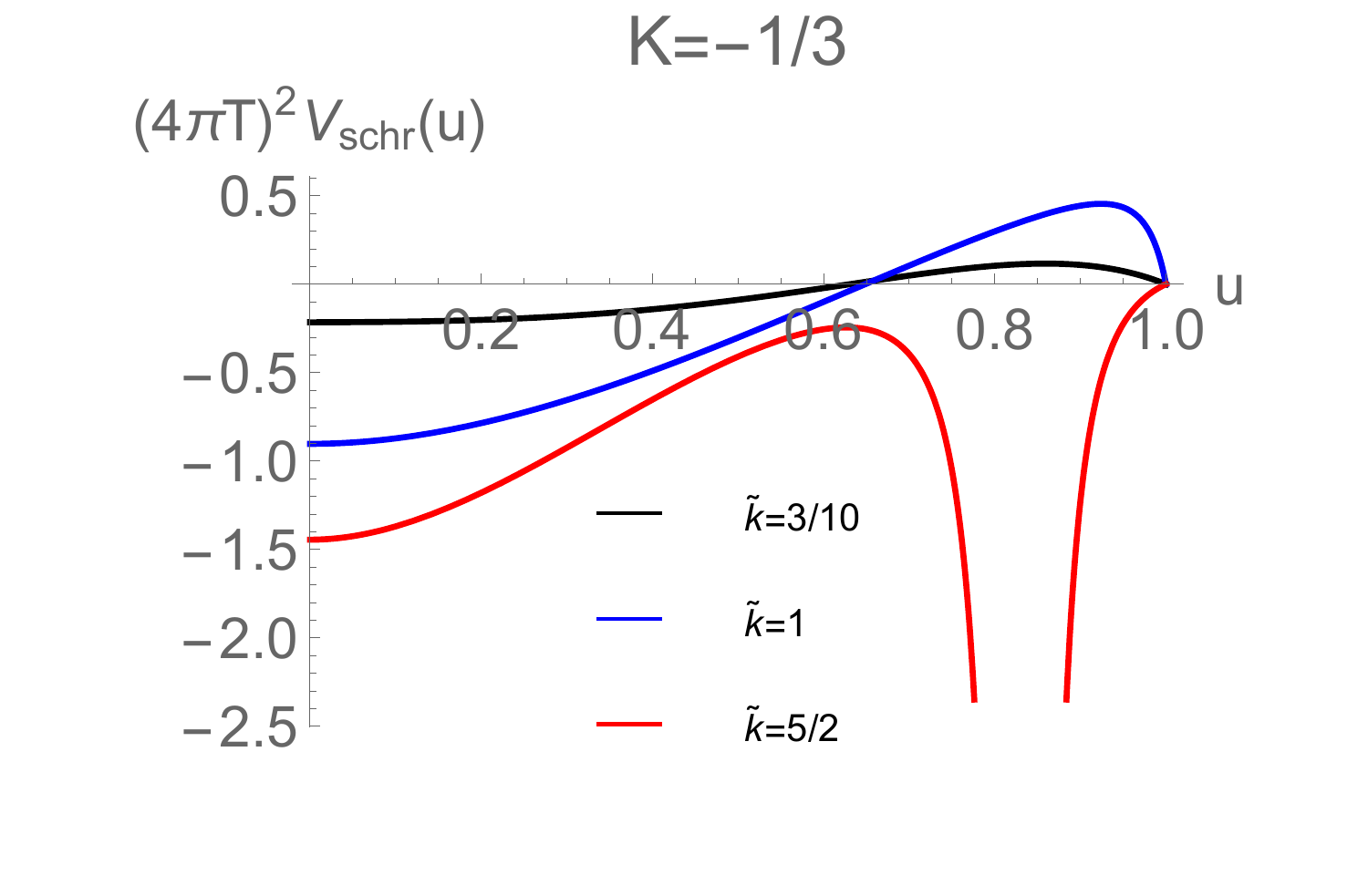}&
\includegraphics[width=0.47\textwidth]{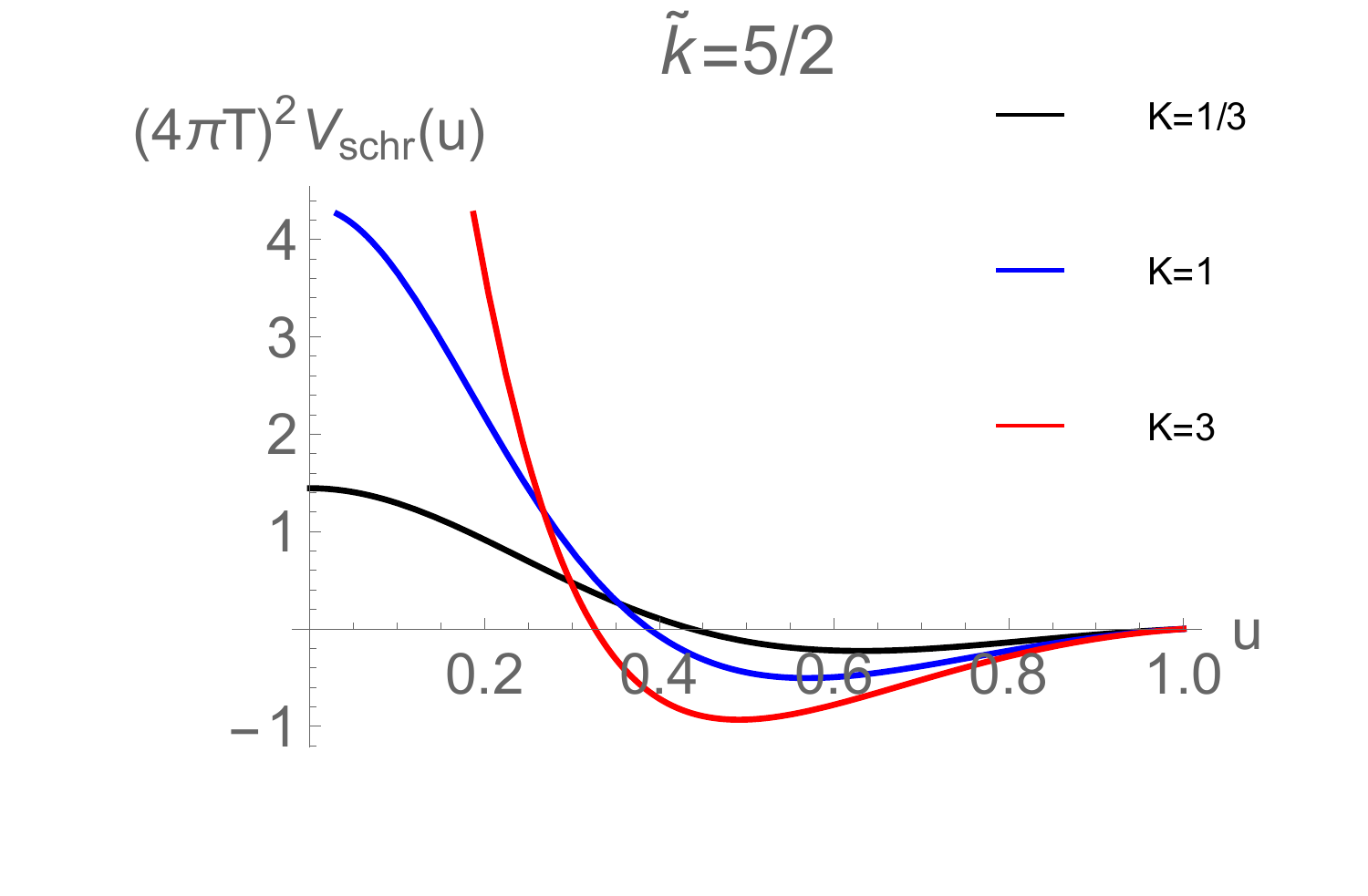}
\end{tabular}
\caption{Plots of the Schr\"odinger potential \eqref{Vschrcase3} versus the radial coordinate $0<u<1$. }
\label{fig:SchroedPot3vskrho=0}
\end{figure}

The perturbation equation reads
\begin{equation}
\label{eq:axeomKrho=0}
\left[\left(1+\frac{\mathcal{K}k^2}{r^2}\right) D a_x'\right]'+\left(1+\frac{\mathcal{K}k^2}{r^2}\right)\frac{\omega^2}{D} a_x=0
\end{equation}
 Accordingly, the Schr\"odinger potential is
\begin{equation}
\label{Vschrcase3}
V_{schr}(r)=\frac{k ^2 \mathcal KD(r) \left[D(r) \left(2 k ^2 \mathcal K+3 r^2\right)-r D'(r) \left(k^2 \mathcal K+r^2\right)\right]}{\left(k^2 \mathcal K r+r^3\right)^2}\,.
\end{equation}
We plot it for relevant values of $\mathcal K$ in figure \ref{fig:SchroedPot3vskrho=0}. The same features as before are seen, with negative wells developing between the boundary and the horizon for positive $\mathcal K$. They are deepest at $T=0$ and for $\mathcal K$ large. We have checked with \eqref{BoundStateFormation} that the area of the negative well is a monotonously increasing function of $\mathcal K$ and accommodates the first zero energy bound state for $\mathcal K\gtrsim3/10$.

Small negative $\mathcal K>-1/6$ shows the appearance of a maximum close to the horizon. There is also a negative well close to the boundary, which is deepest for large $\tilde k$ and smaller values of $\mathcal K$. We have checked that it cannot get big enough for a bound state to form using \eqref{BoundStateFormation}. However, at $\mathcal K\leq-1/6$, an infinite negative well appears for large enough values of $k$, signalling an instability.

To summarize, the allowed values of $\mathcal K$ from the zero density Schr\"odinger potential analysis are $-1/6<\mathcal K\lesssim 3/10$.

\bibliographystyle{JHEP}
\bibliography{AxionsGeneral}

\end{document}